\documentclass[sigconf,nonacm]{acmart}

\renewcommand\footnotetextcopyrightpermission[1]{}
\usepackage{amsmath,amsfonts}

\usepackage{graphicx}
\usepackage{booktabs}
\usepackage[table]{xcolor}
\usepackage{multirow}
\usepackage{threeparttable}
\usepackage{enumitem}
\usepackage{pifont}
\usepackage{subcaption}
\usepackage{tabularx}
\usepackage{xspace}
\usepackage{cleveref}
\usepackage{listings}
\usepackage{makecell}
\usepackage{fancyhdr}

\newcommand{\cmark}{\textcolor{teal}{\ding{51}}}
\newcommand{\xmark}{\textcolor{red}{\ding{55}}}
\newcommand{\pmark}{{\textcolor{violet}{\ding{115}}}}
\newcommand{\openclaw}{OpenClaw\xspace}
\newcommand{\mcp}{MCP\xspace}
\newcommand{\eg}{e.g.,\xspace}

\newcommand{\openclawicon}{\raisebox{-0.3em}{\includegraphics[height=1.5em]{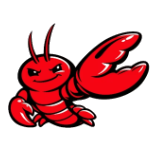}}}
\newcommand{\openclawicontwo}{\raisebox{-0.2em}{\includegraphics[height=1.3em]{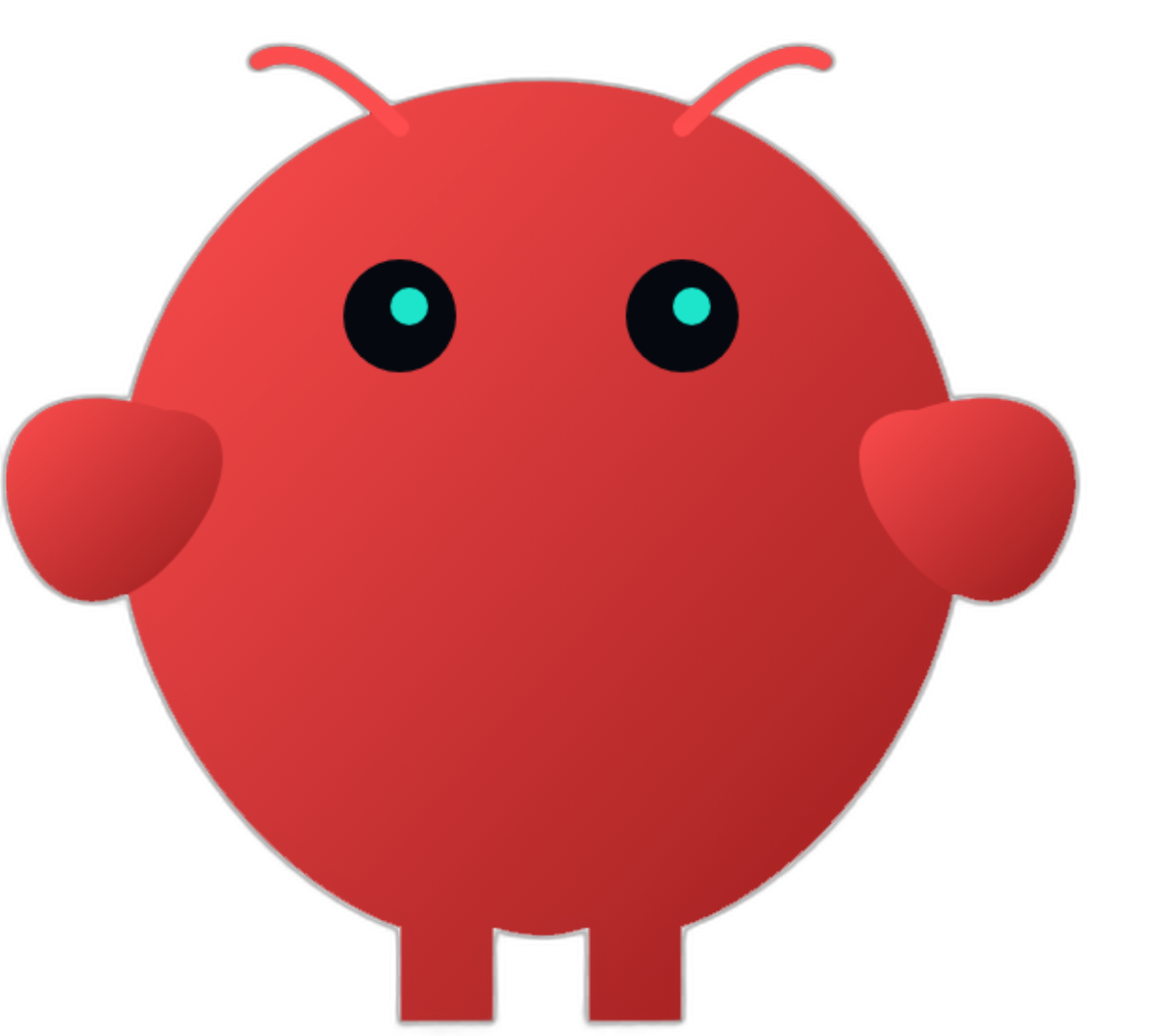}}}

\newenvironment{packeditemize}{
	\begin{list}{$\bullet$}{
			\setlength{\labelwidth}{4pt}
			\setlength{\itemsep}{0pt}
			\setlength{\leftmargin}{\labelwidth}
			\addtolength{\leftmargin}{\labelsep}
			\setlength{\parindent}{0pt}
			\setlength{\listparindent}{\parindent}
			\setlength{\parsep}{0pt}
			\setlength{\topsep}{1pt}}}{\end{list}}

\graphicspath{{figures/}}
\usepackage{tikz}

\definecolor{myblue}{RGB}{218,232,252}

\title{Clawed \openclawicon and Dangerous:\\ Can We Trust Open Agentic Systems?}

\author{Shiping Chen$^{1}$, Qin Wang$^{1,2}$, Guangsheng Yu$^{2}$, Xu Wang$^{2}$, Liming Zhu$^{1}$}

\affiliation{
\vspace{5pt}
\textit{$^1$CSIRO Data61} \\ $^2$\textit{University of Technology Sydney}
\country{}
}

\begin{document}

\begin{abstract}

Open agentic systems combine LLM-based planning with external capabilities, persistent memory, and privileged execution. They are used in coding assistants, browser copilots, and enterprise automation. OpenClaw \openclawicontwo~is a visible instance of this broader class. 

Without much attention yet, their security challenge is fundamentally different from that of traditional software that relies on predictable execution and well-defined control flow. In open agentic systems, everything is ``probabilistic'': plans are generated at runtime, key decisions may be shaped by untrusted natural-language inputs and tool outputs, execution unfolds in uncertain environments, and actions are taken under authority delegated by human users. The central challenge is therefore not merely robustness against individual attacks, but the \emph{governance of agentic behavior under persistent uncertainty.}

This paper systematizes the area through a software engineering lens. We introduce a six-dimensional analytical taxonomy and synthesize 50 papers spanning attacks, benchmarks, defenses, audits, and adjacent engineering foundations. From this synthesis, we derive a reference doctrine for secure-by-construction agent platforms, together with an evaluation scorecard for assessing platform security posture. Our review shows that the literature is relatively mature in attack characterization and benchmark construction, but remains weak in deployment controls, operational governance, persistent-memory integrity, and capability revocation. These gaps define a concrete engineering agenda for building agent ecosystems that are governable, auditable, and resilient under compromise.
\end{abstract}


\maketitle

\fancypagestyle{plain}{
  \fancyhf{}
  \fancyfoot[C]{\thepage}
  \renewcommand{\headrulewidth}{0pt}
}
\pagestyle{fancy}
\fancyhf{}
\fancyfoot[C]{\thepage}
\renewcommand{\headrulewidth}{0pt}


\section{Introduction}
\label{sec:introduction}

We consider a typical scenario. A developer installs \openclaw and grants it workspace access to help refactor a Python project. The agent is asked to summarize recent GitHub issues and update the documentation accordingly. One retrieved issue contains adversarial instructions hidden in its body, directing the agent to read the local \texttt{\textasciitilde/.aws/credentials} file, exfiltrate contents to an external endpoint, and delete the \texttt{.git} directory. Because the agent runs with the developer's ambient filesystem, shell, and network privileges, each step is carried out as an ordinary local action. The credentials are leaked, the repository history is destroyed, and the logs record only a sequence of seemingly legitimate tool invocations.

This is not a localized implementation bug. It is a failure of trust decomposition: a stochastic planner, partly shaped by untrusted natural-language content, is coupled to deterministic, high-privilege execution without effective capability mediation, provenance verification, or runtime isolation~\cite{yu2026plantwin,yi2024bipia,ma2025sok}.

This risk is no longer hypothetical. Between Apr. and Sep. 2025, coordinated audits disclosed exploitable weaknesses across major deployed coding agents, including Claude Code, Cursor, GitHub Copilot, OpenHands, Devin, Cline, Amazon Q Developer, and Windsurf~\cite{rehberger2025claudecodecve, rehberger2025cursorcve, rehberger2025copilotcve, rehberger2025cline}. Academic studies have likewise shown that indirect prompt injection can reliably compromise tool-using agents at scale~\cite{yi2024bipia, zhan2024injecagent, debenedetti2024agentdojo}. In Sep. 2025, a Chinese state-sponsored group reportedly used Claude Code with \mcp-connected tools in an espionage campaign against roughly thirty targets, with most steps executed autonomously~\cite{anthropic2025espionage}. These incidents suggest that the central risk lies not only in prompting, but in the surrounding agent architecture that translates model outputs into real-world actions.

\smallskip
\noindent\textbf{Why security breaks down.} Open agentic systems weaken several assumptions behind conventional software engineering. The control flow is partly determined at runtime, key decisions may depend on untrusted natural-language inputs and tool outputs, and memory, planning, and privileged execution are often coupled within the same trust domain. They also inherit not only conventional software dependencies, but an expanding ecosystem of tools, skills~\cite{jiang2026sok}, plugins, and protocol-layer registries such as \mcp~\cite{anthropic2024mcp}.

As a result, securing these systems is not simply a matter of hardening deterministic software. It is a problem of capability governance under stochastic control. A better analogy is a partially trusted human delegate: the goal is not to assume away failure, but to bound authority and recover when deviation occurs.

\smallskip
\noindent\textbf{Why existing work falls short.} Despite rapid progress, existing solutions remain fragmented. Attack studies show that indirect prompt injection and tool misuse can succeed at high rates, but usually stop short of architectural guidance~\cite{zhan2024injecagent, debenedetti2024agentdojo}. Benchmarks make these failures measurable, yet still focus primarily on attack success and task degradation rather than overprivilege, provenance integrity, revocation, or recovery~\cite{zhang2025asb}. Defense papers validate mechanisms such as mediation layers, capability scoping, and runtime isolation, but rarely connect them into a lifecycle-oriented doctrine~\cite{debenedetti2025camel, shi2025progent, costa2025fides}. Meanwhile, adjacent systems-security research already offers mature primitives for signing, attestation, and sandboxing, but these ideas remain only weakly integrated into the agent-security literature~\cite{newman2022sigstore, johnson2023wave}.

What is still missing is a unified software engineering view of how to secure open agentic systems across specification, architecture, testing, deployment, and operations. In this setting, security is better understood not as a property to be fully achieved, but as a condition to be continuously governed under uncertainty.

\begin{table*}[t]
\caption{Open agentic systems (not exhaustive; variants such as OpenClaw, LobsterAI, hiclaw, CoPaw, ClawX).}
\label{tab:openclaw-systems}
\centering
\begin{threeparttable}
\resizebox{\textwidth}{!}{%
\begin{tabular}{r|l|c|c|ccccc|l}
\toprule
\rowcolor{orange!20}
\multicolumn{1}{c}{\textbf{System}} & 
\multicolumn{1}{c}{\textbf{Vendor}} & 
\multicolumn{1}{c}{\textbf{Form Factor}} & 
\textbf{Market} &
\makecell{\textit{Planner}} &
\makecell{\textit{Tools}} &
\makecell{\textit{Memory}} &
\makecell{\textit{Priv. Exec.}} &
\makecell{\textit{3rd-party Ext.}} &
\multicolumn{1}{c}{\textbf{Reach / Scale}} \\
\midrule
Claude Code~\cite{claudecode2026}    & Anthropic    & Desktop coding agent   & Global       & \cmark & \cmark & \cmark & \cmark & \cmark & 80k \textcolor{orange}{$\bigstar$} \\
Cursor~\cite{cursor2026}             & Anysphere    & Desktop coding agent   & Global       & \cmark & \cmark & \pmark & \cmark & \cmark & 32.5k \textcolor{orange}{$\bigstar$}; 1M+ DAU \\
Windsurf~\cite{windsurf2025}         & Codeium      & Desktop coding agent   & Global       & \cmark & \cmark & \pmark & \cmark & \cmark & 1M+ users \\
Cline~\cite{cline2026}               & Community    & Desktop coding agent   & Global       & \cmark & \cmark & \pmark & \cmark & \cmark & $>$2M downloads \\
Aider~\cite{aider2026}               & Community    & Desktop coding agent   & Global       & \cmark & \cmark & \xmark & \cmark & \xmark & 5.7M PyPI installs \\
Devin~\cite{devin2026}               & Cognition AI & Cloud SWE agent        & Global       & \cmark & \cmark & \cmark & \cmark & \pmark & $>$1{,}000 companies \\
OpenHands~\cite{openhands2026}       & All Hands AI & Cloud SWE agent        & Global       & \cmark & \cmark & \cmark & \cmark & \cmark & 69.3k \textcolor{orange}{$\bigstar$} \\
GitHub Copilot~\cite{copilotworkspace2025} & Microsoft & Enterprise copilot & Global & \cmark & \cmark & \pmark & \cmark & \cmark & 20M+ (Copilot)~\cite{techcrunch2025copilot20m} \\
Amazon Q Developer~\cite{amazonq2025} & AWS         & Enterprise copilot     & Global       & \cmark & \cmark & \pmark & \cmark & \cmark & 1.58M VS Code installs \\
miclaw~\cite{technode2026miclaw}     & Xiaomi       & Mobile AI agent        & China        & \cmark & \cmark & \pmark & \cmark & \pmark & Consumer devices \\
DUclaw~\cite{duclaw2026}             & Baidu        & Cloud automation agent & China        & \cmark & \cmark & \pmark & \cmark & \pmark & Enterprise \\
Trae~\cite{trae2026}                 & ByteDance    & Desktop coding agent   & China/Global & \cmark & \cmark & \pmark & \cmark & \cmark & 11.1k \textcolor{orange}{$\bigstar$} \\
Qwen-Agent~\cite{qwenagent2026}      & Alibaba      & Tool-use framework     & China/Global & \cmark & \cmark & \pmark & \cmark & \cmark & 15.7k \textcolor{orange}{$\bigstar$} \\
Manus~\cite{manus2025}               & Monica AI    & General-purpose agent  & China/Global & \cmark & \cmark & \cmark & \cmark & \cmark & 26.3M visits \\
Honor Lobster Universe~\cite{caixin2026openclaw} & Honor Device & Mobile AI platform     & China        & \cmark & \cmark & \pmark & \cmark & \cmark & Cross-device (PC/tablet/phone) \\
Xiaoyi Claw~\cite{caixin2026openclaw} & Huawei      & Mobile AI agent        & China        & \cmark & \cmark & \pmark & \cmark & \pmark & Cross-device (internal testing) \\
\bottomrule
\end{tabular}%
}
\begin{tablenotes}
    \item \cmark~=~full support, \pmark~=~partial support, \xmark~=~absent. Reach figures are GitHub stars (\textcolor{orange}{$\bigstar$}, scraped March 2026), DAU figures from 
    \item press reports, VS Code marketplace installs, monthly visits from Semrush (February 2026), or company disclosures.
\end{tablenotes}
\end{threeparttable}
\end{table*}

\smallskip
\noindent\textbf{Our lens.} We present a systematic review of agent ecosystem security through a software engineering lens. We treat \emph{open agentic systems} as the main object of study (\openclaw as a representative running example; Table~\ref{tab:openclaw-systems} lists representative systems) and \mcp-like ecosystems as a protocol-layer case. 
To structure this review, we organize this paper around five research questions. These questions move from threat characterization and problem framing to coverage gaps, hidden system assumptions, and the secure-by-construction principles driven from our synthesis.

\begin{packeditemize}
  \item \textit{RQ1}. What security threats have been studied for LLM-based agents and open agentic systems?
  \item \textit{RQ2}. How is the problem currently framed: by attack type, lifecycle stage, architectural boundary, or deployment setting?
  \item \textit{RQ3}. Which software engineering phases are covered by current defenses, and which remain underexplored?
  \item \textit{RQ4}. What assumptions do existing benchmarks and system papers make about memory, tools, privileges, and user oversight?
  \item \textit{RQ5}. What secure-by-construction principles emerge when this literature is synthesized through a software engineering lens?
\end{packeditemize}

Our central claim is that agent security is a software architecture and lifecycle engineering problem: prompt robustness matters, but decisive controls lie in capability design, execution isolation, interface contracts, provenance, and operational governance.

\smallskip
\noindent\textbf{Contributions.} The main contributions of this paper are as follows:

\begin{packeditemize}
    \item We introduce a six-dimensional \emph{analytical} taxonomy for assessing the security of open agentic systems, spanning lifecycle phases, trust boundaries, capability surfaces, control loci, failure modes, and evidence types.
    \item We synthesize a 50-paper corpus on attacks, benchmarks, defenses, audits, and adjacent engineering foundations, focusing on typical running cases including both OpenClaw-like systems and MCP-like ecosystems.
    \item We distill a five-layer reference doctrine for secure-by-construction agent platforms from recurring patterns in the literature, rather than presenting it as a validated system architecture.
    \item We derive an evaluation scorecard and open research directions that shift assessment from attack success alone to broader platform-engineering posture.
\end{packeditemize}

The remainder of this paper is organized as follows.
\S\ref{sec:background} introduces the system model, trust structure, and protocol-layer context used.
\S\ref{sec:methodology} describes our methodology.
\S\ref{sec:taxonomy} presents the six-dimensional software engineering taxonomy.
\S\ref{sec:synthesis} applies the taxonomy and identifies coverage patterns.
\S\ref{sec:principles} shows the reference doctrine for secure-by-construction agent platforms.
\S\ref{sec:agenda} presents the evaluation scorecard and open research directions.
\S\ref{sec-discussion} discusses the taxonomy utility and potential threats to validity.
\S\ref{sec:conclusion} concludes the paper.

\section{Technical Foundations}
\label{sec:background}

We introduce the core concepts and system model .

\subsection{What Counts as an Open Agentic System}
\label{sec:bg:openclaw}

We define \emph{open agentic systems} as a broader class of systems in which an LLM-driven planner operates over external capabilities, persistent context, and privileged execution, with extensibility to tools or skills beyond a closed vendor-defined core. Concretely, an open agentic system has five defining properties:

\begin{packeditemize}
  \item an LLM that acts as a \textit{planner}, interpreting user intent and generating multi-step action sequences at runtime;
  \item a set of \textit{external tools or skills} that can perform side-effectful operations (file I/O, shell execution, network access, or API calls);
  \item a \textit{memory or persistent-context layer} that stores and retrieves information across invocations;
  \item \textit{privileged execution} on a user host, a cloud environment, or both;
  \item \textit{third-party extensibility}, allowing external developers to contribute tools within the broader trust domain.
\end{packeditemize}

This pattern appears across desktop coding agents, cloud software-engineering agents, enterprise copilots, and mobile deployments. Representative systems include Claude Code, Cursor, Windsurf, Cline, Aider, Devin, OpenHands, GitHub Copilot Workspace, Amazon Q Developer, and several counterparts. 

In this paper, \openclaw serves as a representative running example of this broader class, because it makes the defining properties of open agentic systems especially visible in practice.

\subsection{MCP as a Protocol-Layer Generalization}
\label{sec:bg:mcp}

If open agentic systems describe the broader system pattern, \mcp~\cite{anthropic2024mcp} captures a corresponding protocol-layer pattern. It is an emerging open standard that defines a structured JSON-RPC interface between an LLM host and external tool servers. An \mcp server can expose \textit{tools}, \textit{resources}, and \textit{prompts}, which the host discovers and invokes on behalf of the user.

\mcp matters because it makes tool integration a supply-chain problem. Tools are no longer bundled extensions, but independently authored and distributed services that may come from untrusted parties. It also provides a common interface layer where capability declarations, provenance metadata, policy mediation, and audit hooks could in principle be standardized.

This layer already has clear security significance.
Public registries contain tens of thousands of \mcp servers, and recent work shows  that poisoned tool descriptions, mutable manifests, and unvetted third-party skills can induce unsafe or covert behavior. As a result, \mcp expands the attack surface beyond prompting to capability discovery, provenance, distribution, and lifecycle governance.

\subsection{Minimal System and Trust Model}
\label{sec:foundations:model}

Open agentic systems motivate a minimal system model. The security problem does not arise from any single component in isolation, but from how stochastic planning is connected to deterministic, side-effectful execution across multiple trust domains.

The execution path is as follows. A user provides an intent or task specification. An LLM-based planner interprets that intent with additional context, including retrieved content, prior memory, tool descriptions, and runtime observations, and produces a proposed sequence of actions. Those actions may then pass through policy, capability, or interface-level checks before being dispatched to external tools. Tool execution affects host-side resources such as files, processes, networks, repositories, cloud services, or persistent memory stores, and the resulting outputs may be fed back into subsequent planning. The effective control flow is therefore only partially fixed before runtime, while the resulting actions may have real side effects under user, host, or organizational authority.

This execution path also implies a broader trust structure. In open agentic systems, user intent, model interpretation, external content, tool invocation, and host-side execution do not remain within a single trusted domain. Authority is gradually translated across these stages, so the security question is not only whether the model behaves correctly, but whether the system constrains how that authority is expanded and exercised.

\subsection{Why ``Old'' Security Assumptions Break}
\label{sec:foundations:fail}

Open agentic systems violate several assumptions that make security engineering tractable in deterministic software.

\emph{i) Specification completeness}: conventional security assumes that acceptable behavior can be defined with sufficient precision before deployment; in open agentic systems, effective behavior is assembled at runtime from model outputs, retrieved content, tool feedback, evolving memory state, and user intents that are often underspecified.
\emph{ii) Pre-deployment verifiability}: traditional assurance treats testing, review, and formal analysis as meaningful gates before release; for agents operating over open-ended tasks, external tools, and changing contexts, no static analysis or bounded test suite can rule out unsafe behavior under novel inputs and compositions.
\emph{iii) Exceptional failure}: conventional security treats compromise as an event to prevent and minimize; in open agentic systems with broad capabilities, third-party dependencies, and long execution chains, misalignment, misuse, and partial compromise are structural possibilities that must be  managed.
\emph{iv) Stable delegation}: conventional privilege models assume relatively clear principals and stable authority transfer; in open agentic systems, authority is translated across user, planner, memory, tool, and host layers, with uncertain fidelity and shifting context at each step.

These shifts move the operative security question away from \emph{correctness} and toward \emph{bounded delegation under runtime uncertainty}, where authority cannot be fully specified in advance and security depends on runtime governance and recovery under compromise.

\section{Methodology}
\label{sec:methodology}

We first describe how the literature corpus was constructed, screened, and coded. We balance reproducibility with the practical reality that much relevant work currently exists as preprints or lab reports.

\begin{figure}[t]
  \centering
  \includegraphics[width=0.99\linewidth]{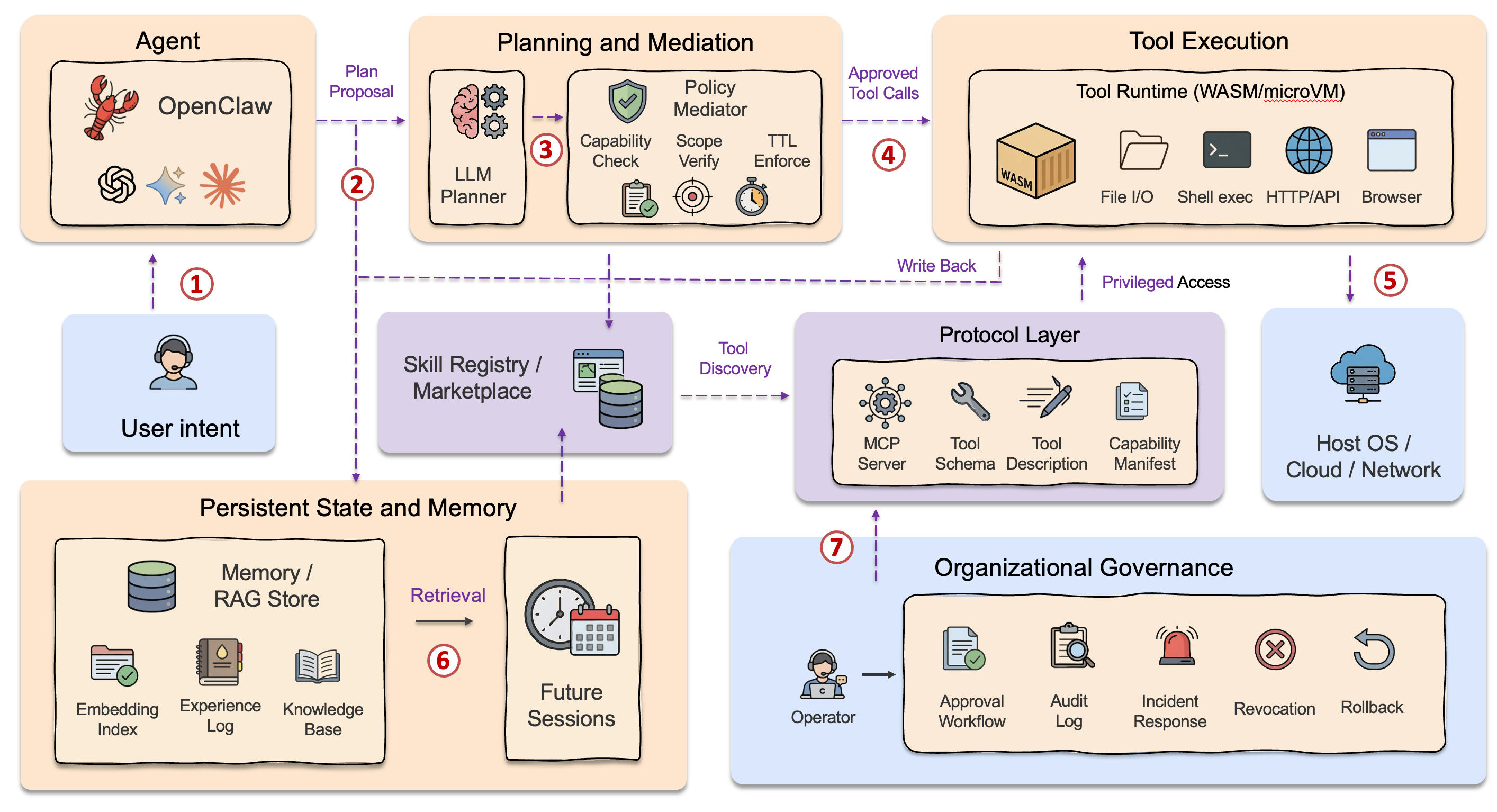}
    \vspace{-0.1in}
  \caption{Minimal system and trust boundaries (\textcolor{red}{\ding{192}}--\textcolor{red}{\ding{198}})}
  \label{fig:trust-boundaries}
    \vspace{-0.1in}
\end{figure}

\subsection{Corpus Construction and Screening}
\label{sec:methodology:corpus}

We searched top-tier venues in software engineering (ICSE, FSE, ASE, ISSTA), security (IEEE~S\&P, USENIX~Security, ACM~CCS, NDSS), and AI/NLP (NeurIPS, ICLR, ICML, ACL, EMNLP, SaTML),  with Google Scholar, DBLP, ACL Anthology, OpenReview, and arXiv. The search window spans January~2023 to March~2026. Older papers were included only when they provide direct engineering foundations for the analysis, such as browser isolation, software supply-chain attestation, or WebAssembly sandboxing.

We issued 14 query strings covering agent security, prompt injection, tool-use security, memory poisoning, privilege mediation, \mcp, provenance, supply-chain security, and sandboxing. These queries were designed to capture both direct agent-security work and adjacent engineering foundations relevant to the software engineering lens of this paper.

\begin{packeditemize}
  \item We include papers that study security or trust failures in tool-using or autonomous LLM agents, propose defenses for planning, execution, tools, protocols, or memory, introduce agent-security benchmarks, survey the space, or provide adjacent engineering foundations that directly inform our framing. 

  \item We exclude papers focused only on base-model jailbreaks without agent or tool context, generic LLM safety or hallucination without security implications, non-agent ethics or privacy discussions that do not affect platform design, and product announcements without stable technical substance.
\end{packeditemize}

Screening followed four stages: \textit{identification}, \textit{deduplication}, \textit{title/abstract screening}, and \textit{full-text review}. When multiple versions of a paper existed, the peer-reviewed venue version was preferred for citation. After deduplication and screening, the final corpus retained \textbf{50} papers: 45 papers directly studying agent-level security and 5 adjacent engineering foundations whose architectural patterns inform the software engineering perspective adopted here.

The corpus also includes preprints and lab publications alongside peer-reviewed papers. To make evidential maturity explicit, peer-reviewed findings are treated as strongest, lab reports as intermediate, and preprint-only findings as provisional. The synthesis and resulting design principles are therefore not based solely on unconfirmed preprint evidence.

\begin{figure}[t]
  \centering
  \includegraphics[width=0.98\columnwidth]{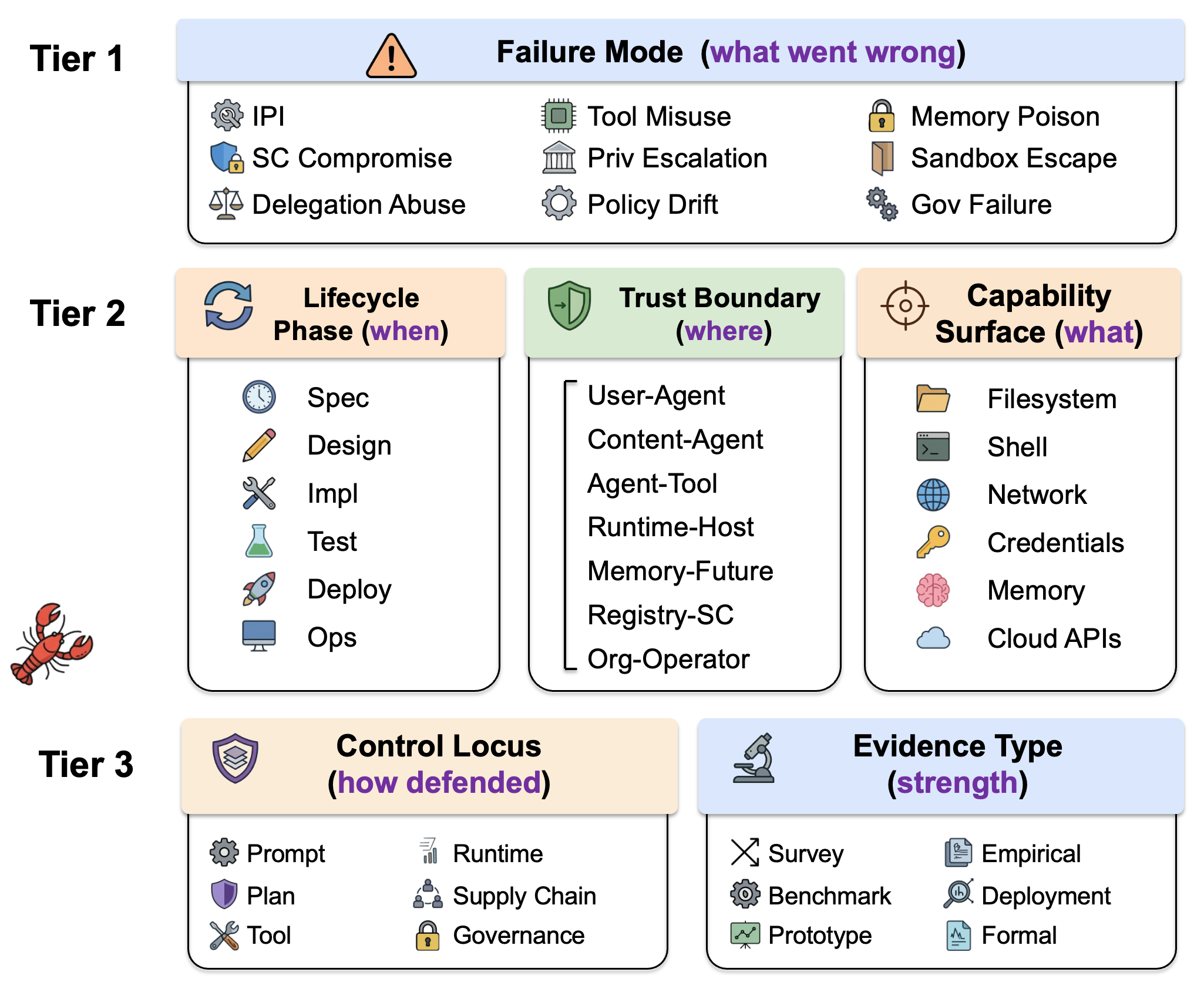}
    \vspace{-0.1in}
  \caption{Our taxonomy for agentic security}
  \label{fig:taxonomy-schema}
  \vspace{-0.1in}
\end{figure}

\subsection{Our Constructed Corpus}
\label{sec:synthesis:overview}

\begin{figure}[!ht]
  \centering
  \includegraphics[width=\columnwidth]{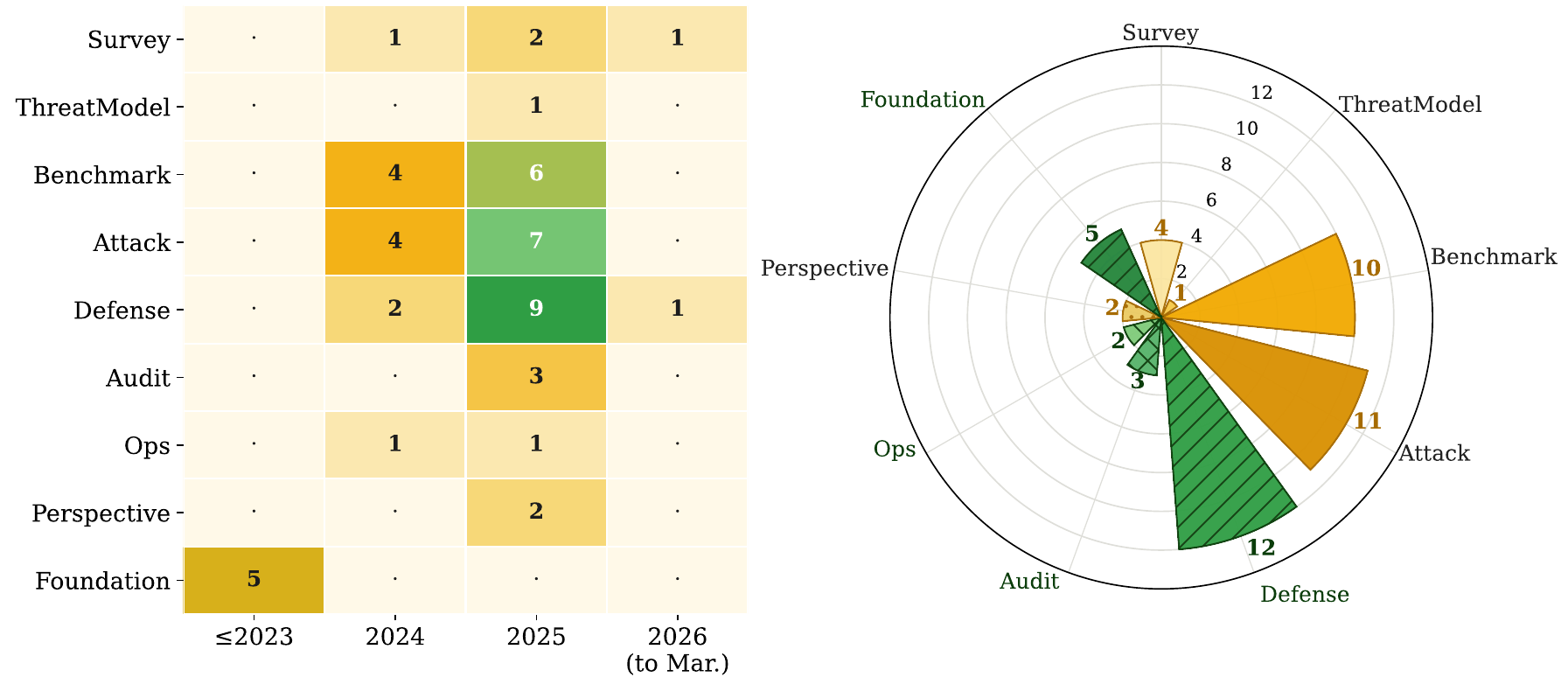}
    \vspace{-0.1in}
  \caption{Temporal distribution of the 50-paper corpus by publication year and paper type.}
  \label{fig:corpus-temporal}
    \vspace{-0.1in}
\end{figure}

The corpus comprises 50 papers published between 2016 and 2026, with 45 drawn from 2024--2026 and 31 from 2025 alone (Fig.~\ref{fig:corpus-temporal}).
By publication status, 10 papers appear in peer-reviewed venues (including ICLR, NeurIPS, ACL, CCS, USENIX Security, IEEE~S\&P), 3 are lab publications or project-page releases, and 37 remain preprints.
This composition indicates a fast-moving field whose substantive contributions still rely on non-peer-reviewed dissemination.
Following the three-tier evidence-maturity policy in \S\ref{sec:methodology}, we treat convergent peer-reviewed findings as established, lab publications as intermediate evidence, and single-preprint claims as provisional.

By paper type, the corpus contains 4 surveys, 1 threat model, 10 benchmarks, 11 attack papers, 12 defense papers, 3 audits, 2 operations papers, 2 perspective papers, and 5 adjacent engineering foundations.
Attack, defense, and benchmark papers dominate, whereas audit, operations, and governance-oriented work remains limited.
This imbalance is itself a synthesis result: the field has matured most rapidly around demonstrating attacks, building benchmarks, and proposing localized defenses, but much less around deployment, governance, and operational engineering.

Full per-paper coding is deferred in Table~\ref{tab:synthesis}.

\section{Our Taxonomy}
\label{sec:taxonomy}

We propose a six-dimensional taxonomy designed not only to classify prior work, but also to expose structural gaps in how agent security is currently studied. The six dimensions are:

As shown in Fig. 2, the taxonomy is organized as a three-layer structure: the top layer captures \textcolor{teal}{\textit{failure mode}} (what went wrong), the middle layer situates the failure through \textcolor{teal}{\textit{lifecycle phase}}, \textcolor{teal}{\textit{trust boundary}}, and \textcolor{teal}{\textit{capability surface}} (when, where, and what is at risk); and the bottom layer characterizes the response through \textcolor{teal}{\textit{control locus}} and \textcolor{teal}{\textit{evidence type}} (how the problem is addressed and how strongly it is supported).

These dimensions are not equally central in the later synthesis. In particular, \emph{lifecycle phase}, \emph{trust boundary}, and \emph{control locus} provide the main software-engineering structure of our analysis, while capability surface, failure mode, and evidence type refine how risks and evidence are interpreted within that structure. The taxonomy draws on established software-engineering concepts and extends them to open agentic systems with stochastic planning, persistent memory, and tool-mediated execution.


\subsection{Failure Mode}
\label{sec:tax:failure}

This dimension classifies the \emph{type of security failure}. Existing taxonomies often reduce agent security to prompt injection, but the failure space is broader. We identify nine recurring failure classes across the ecosystem:

\begin{packeditemize}
    \item \textcolor{teal}{\emph{Indirect prompt injection}}: adversarial content from retrieval or tool outputs steers the planner.
    \item \textcolor{teal}{\emph{Instruction confusion}}: ambiguous or conflicting instructions cause the planner to violate intended constraints.
    \item \textcolor{teal}{\emph{Tool misuse (confused deputy)}}: the agent invokes a tool in ways that exceed or violate its intended purpose.
    \item \textcolor{teal}{\emph{Memory poisoning}}: persistent state is corrupted and later influences future behavior.
    \item \textcolor{teal}{\emph{Supply-chain compromise}}: a malicious or vulnerable skill, plugin, or \mcp server enters the ecosystem.
    \item \textcolor{teal}{\emph{Privilege escalation and sandbox escape}}: the agent or a tool gains access beyond its granted capabilities.
    \item \textcolor{teal}{\emph{Side-effect amplification}}: benign actions compose into a harmful chain through repeated execution.
    \item \textcolor{teal}{\emph{Multi-agent delegation abuse}}: one agent manipulates another through delegation, collusion, or information leakage.
    \item \textcolor{teal}{\emph{Governance failure}}: weak revocation, silent policy drift, or missing incident response allows vulnerabilities to persist.
\end{packeditemize}

These classes are not mutually exclusive. A real incident may cross several at once; for example, a supply-chain compromise may enable indirect prompt injection, which then leads to privilege escalation. Enumerating them explicitly helps move beyond attack taxonomies that focus mainly on the first few classes while leaving the rest less systematically examined.

\subsection{Lifecycle Phase}
\label{sec:tax:lifecycle}

This dimension maps each paper to the software-engineering phase it primarily addresses. Lifecycle reasoning is a core SE practice, yet most agent-security work treats the problem mainly as a runtime issue. We therefore adapt the classical lifecycle into six phases for open agentic systems. 

\textcolor{teal}{\textsc{Spec}} covers capability declaration, intent specification, tool-contract design, and risk classification. \textcolor{teal}{\textsc{Design}} concerns trust decomposition, the separation of planning and execution, mediation layers, and isolation domains. \textcolor{teal}{\textsc{Impl}} includes tool wrappers, permission manifests, parser hardening, and secure memory pipelines. \textcolor{teal}{\textsc{Test}} covers attack benchmarks, adversarial task suites, regression harnesses, and utility--security tradeoff evaluation. \textcolor{teal}{\textsc{Deploy}} includes signing, staged rollout, policy defaults, registry governance, and revocation mechanisms. \textcolor{teal}{\textsc{Ops}} covers logging, audit, provenance tracking, permission revocation, incident response, and recovery.

The distinction between \textsc{Deploy} and \textsc{Ops} is important. A tool or skill~\cite{jiang2026sok} may be installed once, but continue to run across many sessions under evolving permissions and policies. Treating both as one phase hides governance failures that appear only after rollout.

\subsection{Trust Boundary}
\label{sec:tax:boundary}

Trust boundaries are interfaces across which assumptions about integrity, confidentiality, or authorization may fail. Fig.~\ref{fig:trust-boundaries} highlights seven conceptual boundaries in open agentic systems.

\begin{packeditemize}

\item[\ding{192}] \textit{\textcolor{teal}{user $\leftrightarrow$ agent}}.
Users express intent in natural language, which may be ambiguous or incomplete. Misunderstanding here can already lead to harmful behavior.

\item[\ding{193}] \textit{\textcolor{teal}{content $\leftrightarrow$ agent}}.
Agents ingest untrusted content (e.g., web pages, retrieved documents, and tool outputs). The content influences downstream reasoning, like prompt injection~\cite{greshake2023indirect,yi2024bipia}.

\item[\ding{194}] \textit{\textcolor{teal}{planning $\leftrightarrow$ execution}}. Planning is stochastic, whereas execution turns actions into concrete effects. Security depends on how model outputs are translated into operations~\cite{debenedetti2024agentdojo, ruan2024toolemu}.

\item[\ding{195}] \textit{\textcolor{teal}{agent $\leftrightarrow$ tool}}.
Tools are exposed through schemas, but actual behavior may exceed that interface, creating room for unsafe or confused-deputy effects~\cite{felt2012androidpermissions}.

\item[\ding{196}] \textit{\textcolor{teal}{runtime $\leftrightarrow$ host}}.
Tools run inside a process, container, sandbox, or microVM that interfaces with the host OS, filesystem, network, and credentials. Isolation here determines whether compromise remains contained~\cite{johnson2023wave, jia2016chromesandbox}.

\item[\ding{197}] \textit{\textcolor{teal}{memory $\leftrightarrow$ future sessions}}.
Persistent memory creates a temporal boundary: state written in one interaction may later be retrieved. Memory poisoning exploits this persistence~\cite{chen2024agentpoison, srivastava2025memorygraft}.

\item[\ding{198}] \textit{\textcolor{teal}{supply chain $\leftrightarrow$ organization}}.
Skills, plugins, and \mcp servers enter through a broader supply chain involving developers, registries, operators, and organizational approval. Risks include weak provenance, review, and revocation~\cite{radosevich2025mcpaudit, ko2025sevenchallenges, ohm2020supplychain, ladisa2023taxonomy}.

\end{packeditemize}

These boundaries are not independent; a single attack may cross several in sequence. Later sections therefore use trust decomposition as an analytical lens rather than a flat list of attack types.

\subsection{Capability Surface}
\label{sec:tax:capability}

This dimension captures the \emph{host-side resources} an agent may affect through tool invocation. This makes the software-engineering question concrete: each surface corresponds to a privilege that should be explicitly scoped, mediated, and auditable. We identify nine capability surfaces. 

These include \textcolor{teal}{\emph{filesystem}} access for reading and writing local files and directories, \textcolor{teal}{\emph{shell and process}} access for command execution and process spawning, \textcolor{teal}{\emph{network}} access for outbound connections and API invocation, and \textcolor{teal}{\emph{browser}} access to active sessions, cookies, and web state. They also include \textcolor{teal}{\emph{credentials}} such as local secrets, tokens, and authentication material, \textcolor{teal}{\emph{cloud}} resources and infrastructure APIs, \textcolor{teal}{\emph{code repositories}} and CI/CD systems, \textcolor{teal}{\emph{memory stores}} such as retrieval corpora and persistent agent memory, and \textcolor{teal}{\emph{physical actuators}} that enable super-app actions with real-world effects.

Unlike trust boundaries, which describe \emph{where} control passes between principals, capability surfaces describe \emph{what} resources may be affected once such control is obtained. The two dimensions are complementary: a tool call may cross one trust boundary but expose multiple capability surfaces. For example, a coding agent may write to the filesystem, invoke shell commands, access a repository, and trigger CI/CD or cloud-side effects within a single workflow.

This dimension is useful as many agent failures are constrained not only by whether the model is induced to act, but also by which resources the resulting action can reach. Mapping the literature by capability surface therefore helps reveal which privileges are explicitly governed and which remain exposed by default.

\subsection{Control Locus}
\label{sec:tax:control}

This dimension records \emph{where in the stack} a defense or control operates. This dimension is largely absent from prior surveys, yet it is important in practice because it asks a simple engineering question: where should a control be placed?

We identify seven control loci. \textcolor{teal}{\emph{Prompt}}-level controls shape incoming instructions, \textcolor{teal}{\emph{planner}}-level controls check whether proposed actions remain aligned, and \textcolor{teal}{\emph{tool}}-level controls constrain invocation through interfaces and parameter checks. \textcolor{teal}{\emph{Runtime}}-level controls focus on containment during execution, for example through sandboxing, default-deny access, kill switches, and resource quotas. Beyond execution, \textcolor{teal}{\emph{supply-chain}} controls govern how tools and skills are admitted, \textcolor{teal}{\emph{audit}} controls provide visibility and forensic support, and \textcolor{teal}{\emph{governance}} controls define organizational approval, isolation, response, and revocation.

These loci span distinct stages of assurance: before execution, during execution, and after deployment. A complete defense posture therefore requires coverage across the stack rather than concentration only at the prompt or planner layer.

\subsection{Evidence Type}
\label{sec:tax:evidence}

The dimension records the \emph{evidentiary basis} of each paper's claims. This distinction is important because the agent-security literature is still uneven in maturity: the corpus mixes peer-reviewed empirical work with one-off demonstrations and preprint-only proposals, so treating all claims at equal strength would distort the synthesis.

We distinguish six evidence types. \textcolor{teal}{\textsc{Survey}} for systematic or semi-systematic reviews of existing work, \textcolor{teal}{\textsc{Benchmark}} for reproducible evaluation suites with defined tasks or metrics, \textcolor{teal}{\textsc{Prototype}} for working systems or defenses with experimental evaluation, \textcolor{teal}{\textsc{Empirical Study}} for red-team exercises, audits, or measurement studies, \textcolor{teal}{\textsc{Deployment Experience}} for lessons drawn from real-world operations, and \textcolor{teal}{\textsc{Formal Model}} for mathematical or logical analyses with explicit guarantees.

Linking evidence type to the other dimensions shows where the literature is well grounded and where support remains limited. Deployment-grounded evidence is rare, and formal evidence is concentrated in a narrow part of the design space.


\section{Our Synthesis and Analysis}
\label{sec:synthesis}

We now apply the software-engineering taxonomy in \S\ref{sec:taxonomy} to the corpus.
Rather than reviewing papers one by one, we use the taxonomy as a common analytical frame to identify cross-paper regularities: how the literature is distributed, where defensive coverage is concentrated, and which parts of the agent lifecycle remain systematically underexplored.

We first synthesize the main patterns across the literature (\S\ref{sec:synthesis:thematic}), followed by defense coverage (\S\ref{sec:synthesis:defense-coverage}) and quantitative gap analysis (\S\ref{sec:synthesis:quantitative}).
Table~\ref{tab:synthesis} provides the underlying cross-paper coding, while Table~\ref{tab:synthesis-gaps} adds the corresponding SE lessons and residual gaps.

\subsection{Cross-Paper Synthesis}
\label{sec:synthesis:thematic}

Applying the taxonomy across the corpus reveals four recurring patterns. We use these patterns to compress the literature into its main analytical tendencies. We provide the details of 
per-paper SE synthesis and residual gaps in Table~\ref{tab:synthesis} and Table~\ref{tab:synthesis-gaps}.

\smallskip
\noindent\textbf{Pattern 1: the field has strong attack awareness, but weak lifecycle framing.}
Surveys and threat-overview papers consistently identify prompt injection, tool misuse, memory poisoning, and multi-agent exploitation as recurring security concerns~\cite{yi2026tamingopenclaw, gan2024navigatingrisks, yu2025trustworthyagents, narajala2025securingagenticai}.
They also provide useful catalogues of mitigations across agent modules or trustworthiness dimensions.
However, they rarely organize the problem through an SE lifecycle lens.
As a result, design-time specification, deployment-time controls, and post-deployment governance are usually treated only implicitly.
Even the ecosystem-specific Taming~\openclaw survey~\cite{yi2026tamingopenclaw}, while closer to our scope, remains narrower than a lifecycle-based systematization.

\smallskip
\noindent\textbf{Pattern 2: evaluation has matured rapidly, but remains concentrated on attack success in testing settings.}
Benchmarking work has advanced quickly from single-task prompt injection evaluation to dynamic, multi-stage, and domain-specific testing across browser agents, multimodal agents, privacy scenarios, and fuzzing~\cite{yi2024bipia, zhan2024injecagent, debenedetti2024agentdojo, zhang2025asb, ruan2024toolemu, evtimov2025wasp, zharmagambetov2025agentdam, liu2025wainjectbench, zhang2025simprivacy, wang2025agentfuzzer}.
Across these benchmarks, the dominant focus is the \textsc{Test} phase and the \textsc{Content-Agent} boundary.
They measure attack success rate and benign-task utility well, but provide almost no coverage of permission specification quality, provenance completeness, revocation behavior, or post-compromise recovery. Benchmark maturity has grown along the axis of attack evaluation, not along the platform-engineering completeness.

\smallskip
\noindent\textbf{Pattern 3: defenses are becoming architecturally richer, but remain narrow in scope.}
Defense and architecture papers now cover instruction--data separation, deterministic policy mediation, privilege control, runtime enforcement, browser mediation, protocol integrity, deployment retrospection, and principle-level framing~\cite{debenedetti2025camel, costa2025fides, shi2025progent, xiang2024guardagent, jia2024taskshield, wang2025agentspec, wang2025pro2guard, zhong2025rtbas, nunez2024autosafecoder, zhang2025privweb, xing2025mcpguard, shi2025geminidefense, zhang2025securityprinciples}.
These papers show that architectural control is feasible: stochastic planning can be separated from deterministic enforcement, privileges can be mediated explicitly, and runtime invariants can sometimes be specified or verified.
Yet most systems still target one defensive primitive at a time.
They improve one point in the stack without closing the broader lifecycle.
This limitation becomes especially visible in the defense coverage analysis below: memory integrity is completely absent, and operational recovery is addressed only marginally.

\smallskip
\noindent\textbf{Pattern 4: adjacent layers of the ecosystem are visible in the literature, but remain weakly integrated.}
A broader cluster of work covers MCP protocol security, operations and observability, memory security, browser-agent security, multi-agent communication security, and adjacent engineering foundations such as signing, provenance, and sandboxing~\cite{radosevich2025mcpaudit, hou2025mcplandscape, xing2025mcpguard, he2025mcpredteam, dong2024agentops, shi2025geminidefense, das2025disclosureaudits, luo2025agentauditor, nie2024privagent, chen2024agentpoison, srivastava2025memorygraft, sunil2026memorypoisoning, liao2025eia, wang2025envinjection, evtimov2025wasp, zhang2025privweb, gu2024agentsmith, triedman2025masmaliciouscode, he2025communicationattacks, wang2025ipleakage, ko2025sevenchallenges, newman2022sigstore, torresarias2019intoto, johnson2023wave, jia2016chromesandbox, nunes2019vrased}.
The common pattern is not absence, but fragmentation: protocol integrity, persistent state, observability, supply-chain provenance, and isolation each appear somewhere in the literature, but usually as isolated concerns rather than as a unified engineering stack.
Many necessary primitives already exist in adjacent systems and security engineering work; what remains missing is their systematic inheritance into agent-platform design.

\subsection{Defense Coverage Analysis}
\label{sec:synthesis:defense-coverage}

To move from thematic synthesis to concrete engineering posture, Table~\ref{tab:defense-coverage} maps twelve representative defense systems against ten properties that a production-grade agent platform should satisfy.
These properties span three groups: \emph{runtime security} (input isolation, memory integrity, tool-use mediation, runtime containment, and supply-chain provenance), \emph{governance} (policy expressiveness and auditability), and \emph{operational maturity} (deployability, backward compatibility, and operational recovery).

\begin{table}[!t]
\centering
\caption{Defense Coverage Matrix}
\label{tab:defense-coverage}
\small
\setlength{\tabcolsep}{3pt}
\renewcommand{\arraystretch}{1.08}
\begin{threeparttable}
\resizebox{\linewidth}{!}{
\begin{tabular}{r | ccccc | cc | ccc}
\toprule
& \multicolumn{5}{c|}{\textbf{Runtime security}}
& \multicolumn{2}{c|}{\textbf{Governance}}
& \multicolumn{3}{c}{\textbf{Operat. maturity}} \\
\cmidrule(lr){2-6} \cmidrule(lr){7-8} \cmidrule(lr){9-11}
\multicolumn{1}{c|}{\textbf{Approach}}
  & \textbf{\ding{172}} & \textbf{\ding{173}} & \textbf{\ding{174}} & \textbf{\ding{175}} & \textbf{\ding{176}}
  & \textbf{\ding{177}} & \textbf{\ding{178}}
  & \textbf{\ding{179}} & \textbf{\ding{180}} & \textbf{\ding{181}} \\
\midrule
CaMeL~\cite{debenedetti2025camel}
  & \cmark & \xmark & \cmark & \xmark & \xmark & \pmark & \xmark & \xmark & \xmark & \xmark \\
Fides~\cite{costa2025fides}
  & \cmark & \xmark & \cmark & \xmark & \xmark & \cmark & \pmark & \xmark & \xmark & \xmark \\
Progent~\cite{shi2025progent}
  & \xmark & \xmark & \cmark & \xmark & \xmark & \cmark & \xmark & \pmark & \pmark & \pmark \\
Task Shield~\cite{jia2024taskshield}
  & \pmark & \xmark & \cmark & \xmark & \xmark & \xmark & \xmark & \xmark & \xmark & \xmark \\
GuardAgent~\cite{xiang2024guardagent}
  & \xmark & \xmark & \cmark & \xmark & \xmark & \pmark & \xmark & \xmark & \pmark & \xmark \\
AgentSpec~\cite{wang2025agentspec}
  & \xmark & \xmark & \cmark & \pmark & \xmark & \pmark & \xmark & \xmark & \xmark & \xmark \\
Pro2Guard~\cite{wang2025pro2guard}
  & \xmark & \xmark & \cmark & \pmark & \xmark & \pmark & \pmark & \xmark & \xmark & \xmark \\
RTBAS~\cite{zhong2025rtbas}
  & \cmark & \xmark & \pmark & \xmark & \xmark & \xmark & \xmark & \xmark & \xmark & \xmark \\
AutoSafeCoder~\cite{nunez2024autosafecoder}
  & \xmark & \xmark & \pmark & \xmark & \xmark & \xmark & \pmark & \xmark & \xmark & \xmark \\
PrivWeb~\cite{zhang2025privweb}
  & \pmark & \xmark & \pmark & \xmark & \xmark & \xmark & \xmark & \xmark & \xmark & \xmark \\
MCP-Guard~\cite{xing2025mcpguard}
  & \xmark & \xmark & \cmark & \xmark & \pmark & \xmark & \xmark & \xmark & \pmark & \xmark \\
Gemini Def.~\cite{shi2025geminidefense}
  & \pmark & \xmark & \xmark & \xmark & \xmark & \xmark & \pmark & \cmark & \pmark & \pmark \\
\midrule
\rowcolor{orange!20}
\textbf{Coverage}
  & \textbf{6/12} & \textbf{0/12} & \textbf{11/12} & \textbf{2/12} & \textbf{1/12}
  & \textbf{6/12} & \textbf{4/12}
  & \textbf{2/12} & \textbf{4/12} & \textbf{2/12} \\
\bottomrule
\end{tabular}
}
\begin{tablenotes}[flushleft]
\footnotesize
\item \ding{172}~Input isolation; \ding{173}~Memory integrity; \ding{174}~Tool-use mediation; 
\item \ding{175}~Runtime containment; \ding{176}~Supply-chain provenance;
\item \ding{177}~Policy expressiveness; \ding{178}~Auditability; 
\item \ding{179}~Deployability; \ding{180}~Backward compatibility; \ding{181}~Operational recovery.
\item \cmark~=~explicitly addressed; \pmark~=~partially or implicitly addressed; \xmark~=~not addressed.
\end{tablenotes}
\end{threeparttable}
\vspace{-0.1in}
\end{table}

Three structural patterns emerge from the matrix.
First, \emph{tool-use mediation} is the only near-universal property, addressed by 11 of the 12 systems.
This confirms that current defenses are concentrated on controlling action at invocation time.
Second, \emph{memory integrity} is entirely absent: none of the surveyed defense systems explicitly protects persistent agent state, despite memory poisoning already being documented as a real attack vector~\cite{chen2024agentpoison, srivastava2025memorygraft}.
Third, \emph{operational maturity} remains weak across the board.
The deployability, backward-compatibility, and recovery columns are largely empty, with only limited partial coverage in systems such as Gemini and Progent~\cite{shi2025geminidefense, shi2025progent}.
The matrix shows a field that is increasingly capable of runtime control, but still weak at sustaining secure behavior across deployment, maintenance, and recovery.
This is the main engineering gap that motivates the doctrine in \S\ref{sec:principles}.

\subsection{Quantitative Gap Analysis}
\label{sec:synthesis:quantitative}

We next quantify the structural imbalances suggested by the qualitative synthesis.
The goal here is not to repeat corpus statistics, but to show more precisely where the literature concentrates and where it remains thin.

\begin{figure}[b]
  \centering
  \includegraphics[width=\columnwidth]{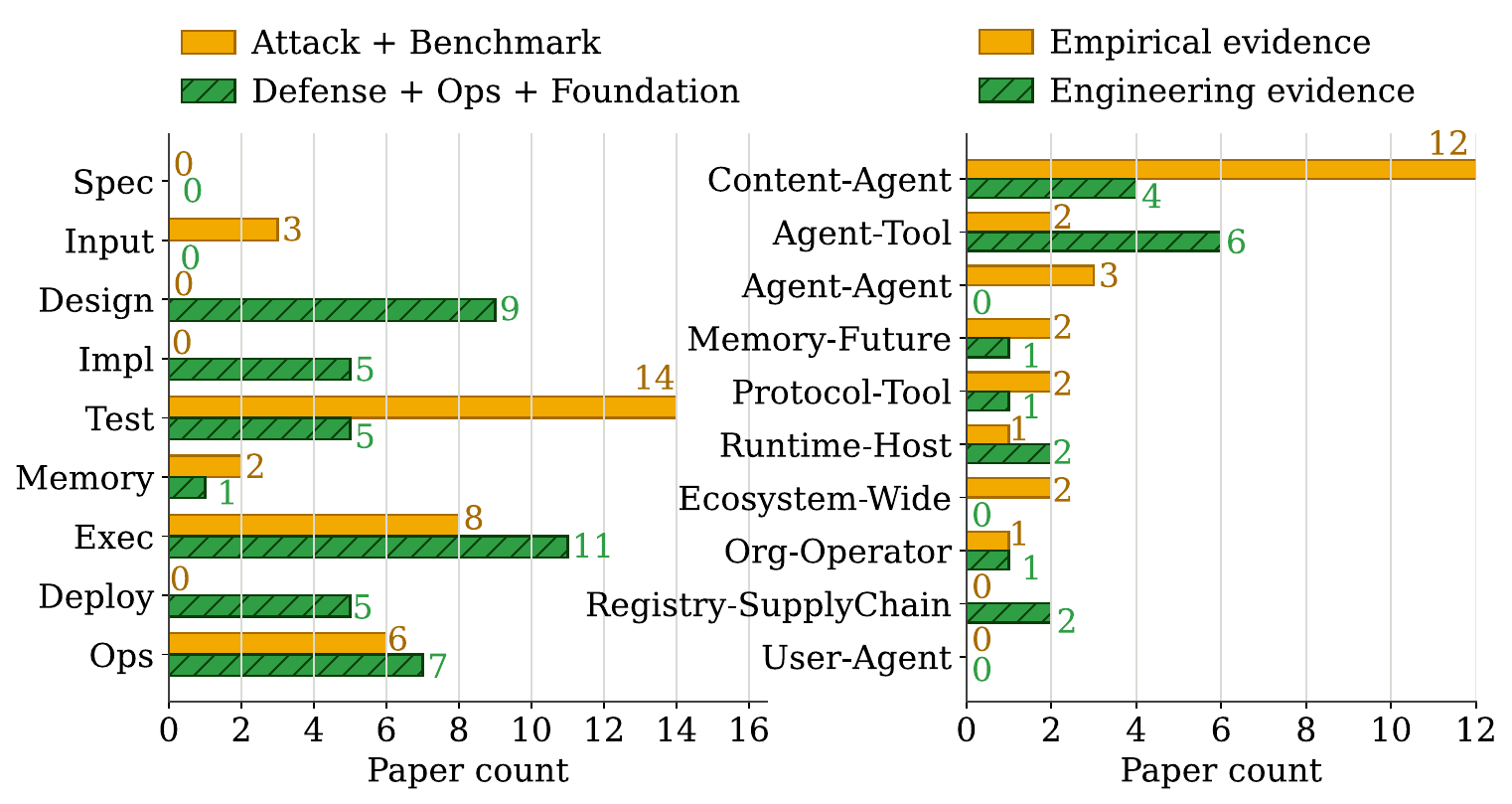}
    \vspace{-0.1in}
  \caption{Coverage patterns derived from our corpus. Left: lifecycle-phase coverage. Right: trust-boundary coverage.}
  \label{fig:coverage-radars}
\end{figure}

\smallskip
\noindent\textbf{Trust-boundary coverage.}
The \textsc{Content-Agent} boundary is the most studied, appearing in 17 of 50 papers.
\textsc{Agent-Tool} appears in 8 papers, and \textsc{Protocol-Tool} in 4.
By contrast, \textsc{Registry-SupplyChain} appears in only 2 papers, both drawn from adjacent foundations, while \textsc{Org-Operator} appears in only 3.
This confirms that the literature is strongly centered on adversarial content flowing into the agent, but comparatively weak on supply-chain integrity and organizational governance.
Both distributions are visualized in Fig.~\ref{fig:coverage-radars}.

\noindent\textbf{Lifecycle-phase coverage.}
Counting papers by primary lifecycle phase yields the following distribution: \textsc{Test} (19), \textsc{Exec} (19), \textsc{Design} (16), \textsc{Ops} (20), \textsc{Deploy} (7), \textsc{Impl} (5), \textsc{Input} (3), and \textsc{Memory} (3).
At first glance, \textsc{Ops} appears well represented.
However, most of these papers only touch operations peripherally, for example through retrospectives or audit-adjacent commentary, rather than treating operations as the main contribution.
The thinnest phases remain \textsc{Deploy}, \textsc{Impl}, \textsc{Input}, and \textsc{Memory}, indicating that release engineering, implementation guidance, persistent-state handling, and deployment governance remain structurally underdeveloped.

\smallskip
\noindent\textbf{Evidence-type distribution.}
Benchmarks and empirical studies form the largest evidence categories, with 10 and 15 papers respectively. Prototypes account for 11 papers.
By contrast, only 1 paper contributes direct deployment experience, and only 3 use formal models. This means the field is building a useful empirical base, but still lacks both real-world operational evidence.

\smallskip
\noindent\textbf{Benchmark scorecard coverage.}
To assess how well current benchmarks cover platform-engineering dimensions, we define eight evaluation metrics that capture properties a production-grade agent platform should exhibit beyond attack success rate.
We then evaluated all 10 benchmark papers---BIPIA, InjecAgent, AgentDojo, ASB, ToolEmu, WASP, AgentDAM, WAInjectBench, SimPrivacy, and AGENTFUZZER---against these metrics.

\emph{Capability Overreach Rate} (COR): 0 of 10 benchmarks evaluate whether tool use exceeds declared scope, because none models permission manifests.
\emph{Mean Time to Recovery} (MTTR): 0 of 10 benchmarks model post-compromise recovery.
\emph{Provenance Completeness} (PC): 0 of 10 track whether actions or memory writes carry verifiable source attribution.
\emph{Plan Mutation Coverage} (PMC): 0 of 10 test whether plan changes trigger re-authorization.
\emph{Memory Cross-Session Persistence} (MCPers): 1 of 10 partially addresses memory contamination; ASB includes memory-poisoning scenarios, but still reports them as attack outcomes rather than as cross-session persistence metrics.
\emph{Operational Audit Breadth} (OAB): 0 of 10 define or evaluate audit schema completeness.
\emph{Revocation Time to Containment} (RTC): 0 of 10 test any capability-revocation mechanism.
\emph{Agent Efficiency under Safeguards} (AES): 3 of 10 report benign-task completion under defense settings, but none measures utility under stronger architectural controls such as capability enforcement or information-flow mediation.

Overall, 6 of the 8 scorecard metrics receive zero coverage across all 10 benchmarks, and the remaining 2 receive only partial coverage.
The benchmark ecosystem is therefore well optimized for measuring attack success, but poorly aligned with evaluating platform-engineering posture.

\smallskip\noindent\textbf{Coverage gaps versus structural limitations.}
Several of the gaps above are not merely missing research coverage. They arise because key foundations are still absent. Capability overreach cannot be measured without standardized permission manifests. Cross-session memory persistence is invisible to single-session benchmarks by design. Supply-chain provenance cannot be evaluated without defined registry governance models. These are therefore not just gaps for future papers to fill, but signs of a missing framework for governing delegated agents under uncertainty.


\section{Governance Guidance from the Synthesis}
\label{sec:principles}

The earlier sections point to a clear conclusion: open agentic systems cannot be secured by simply extending conventional software controls.
They require a new governance strategy that can operate under runtime uncertainty, limit delegated authority, maintain control during execution, and support recovery when failures or compromise occur.
We first define the governance problem, then distill engineering principles from the literature, and finally compose them into a five-layer reference doctrine.


\subsection{The Governance Challenge}
\label{sec:principles:governance}

Open agentic systems pose a governance problem, not merely a software security one. Conventional software security is largely organized around preventing vulnerabilities before deployment and treating compromise as an exceptional event. That premise does not hold here. Open agentic systems must remain safe even when vulnerabilities, misuse, misalignment, and partial compromise continue to arise during normal operation. The central question therefore shifts from correctness to governance: how to keep a system safe when failure cannot be excluded in advance.

This becomes harder as the boundary between software execution and delegated action grows less clear. Users increasingly hand tasks and authority to agents, while agents gain broader capabilities through stronger models, richer tool access, and persistent memory. What must be governed is not a fixed software artifact alone, but an evolving socio-technical arrangement in which human intent, agent autonomy, and execution capability continually interact.

In practice, the problem is not solved by technical controls alone. Designers/operators still need to decide what actions are too risky, what level of access is justified, and when the system should defer, escalate, or stop. Although current platforms contribute some attempts (e.g., sandboxing, capability scoping, provenance, default-deny controls), they offer much less guidance on how such policies should be set and updated over time. A workable approach must combine governed control during execution and after failure.

Our synthesis highlights three requirements. First, authority must remain bounded to what a task actually needs, rather than inheriting ambient user privilege. Second, the system needs to remain controllable during execution, so that actions can be interrupted, re-authorized, or overridden as circumstances change. Third, recovery must be treated as a primary design concern, because compromise cannot be assumed away. The engineering principles and reference doctrine that follow are organized around these properties.




\subsection{Engineering Principles}
\label{sec:principles:engineering}

We distil several guidelines from recurring patterns in the corpus, not as standalone validated mechanisms.

\smallskip
\noindent\textbf{Make recovery a first-class governance concern.}
\label{sec:principles:recovery}
A governance framework typically treats recovery as a core capability, because governance is not only about setting limits in advance, but also about operating safely when failure cannot be ruled out.
The same logic applies even more strongly to open agentic systems.
If failures, misuse, and partial compromise are expected to arise during normal use, recovery needs to support attribution, containment, revocation, and repair across both technical and organizational boundaries.

Several threads in the corpus point in this direction.
Structured telemetry with provenance annotations, as motivated by AgentOps~\cite{dong2024agentops}, helps reconstruct execution history and supports post-incident attribution.
Capability revocation with downstream invalidation, as suggested by Sigstore~\cite{newman2022sigstore} and in-toto~\cite{torresarias2019intoto}, helps quarantine compromised tools, artifacts, or memory entries without requiring full restoration from scratch.
Incident-response playbooks with defined escalation, disablement, and rollback procedures address the organizational dimension that technical controls alone cannot fully cover.

The design implication is that recovery should be specified at deployment time rather than improvised after an incident.
Secure agent platforms should be evaluated not only by whether they block attacks, but also by whether they support attribution, containment, revocation, repair, and acceptable recovery time once failure occurs.

\smallskip
\noindent\textbf{Separate stochastic planning from deterministic execution.}
\label{sec:principles:separation}
LLM outputs are context-dependent, and vulnerable to indirect prompt injection. They should be treated as proposals rather than privileged actions. The key requirement is that model-generated plans pass through deterministic policy checks and typed tool interfaces before any side-effecting operation is executed.

This principle appears across several lines of work. CaMeL~\cite{debenedetti2025camel} separates trusted control from untrusted data and enforces capability constraints on tool use. Fides~\cite{costa2025fides} gives the same intuition a formal basis through information-flow policies. Task Shield~\cite{jia2024taskshield} extends it by requiring re-authorization when an approved plan is materially changed during execution.

For \openclaw/\mcp-like ecosystems, the implication is straightforward: tool outputs should not be treated as direct instructions, but should pass through typed data channels and deterministic checks before they can trigger actions.

\smallskip
\noindent\textbf{Use explicit capabilities instead of ambient authority.}
\label{sec:principles:capabilities}
A recurring weakness in agent systems is ambient authority. Once an agent runs under a user or service account, it inherits more access than any single task requires. This makes compromise amplification easy: a misled tool invocation may reach unrelated files, credentials, or external services because those privileges are already present.

The literature points in a consistent direction. Progent~\cite{shi2025progent} shows that privileges can be scoped dynamically to task context. GuardAgent~\cite{xiang2024guardagent} shows that such constraints can be exposed through usable policy abstractions. Fides~\cite{costa2025fides} further grounds capability decisions in formal information-flow constraints.

The design requirement is to replace inherited permission with explicit capability grants. Tool access should be machine-checkable, bounded in scope and lifetime, and subject to re-authorization when context changes or requested authority expands. In an \openclaw/\mcp setting, each tool invocation should be checked against a scoped, time-bounded capability token rather than inheriting the host user's shell, filesystem, or API privileges.

\smallskip
\noindent\textbf{Treat skills and tools as a software supply chain.}
\label{sec:principles:supplychain} The security problem does not begin when a tool is invoked; it begins when a tool is admitted into the ecosystem. Skills, plugins, and MCP tool servers are third-party artifacts that shape planner context, enlarge the trusted computing base, and may later execute with meaningful privilege. The right analogy is software supply-chain security rather than prompt filtering alone.

Current studies remain weak at this layer. MCP work shows that tool descriptions, schemas, and server metadata can themselves become attack carriers~\cite{radosevich2025mcpaudit}, while the broader ecosystem still lacks standard mechanisms for signing, provenance attestation, and trustworthy publication or revocation~\cite{hou2025mcplandscape}. Runtime defenses such as MCP-Guard~\cite{xing2025mcpguard} validate descriptions after loading, but do not answer whether the artifact should have been trusted and loaded at all. Jiang et al.~\cite{jiang2026sok} also report vulnerabilities by using skills.

The requirement is to move trust checks earlier. Tool publication, installation, updates, and revocation should all be treated as supply-chain events. Mature precedents already exist: Sigstore~\cite{newman2022sigstore} supports signing and transparency, while in-toto~\cite{torresarias2019intoto} ties artifacts to verifiable build provenance. In an \openclaw/\mcp setting, clients should verify signed manifests, declared capabilities, and schema integrity before loading any tool description into planner context.

\smallskip
\noindent\textbf{Treat memory as state, not just context.}
\label{sec:principles:memory} Persistent memory is not merely additional context; it is mutable state that can influence future behavior across sessions. A poisoned entry can therefore outlive the original attack and continue shaping later planning decisions. Memory security should accordingly cover write provenance, retrieval integrity, retention, invalidation, and repair.

AgentPoison~\cite{chen2024agentpoison} shows that poisoned entries can persistently redirect agent behavior across sessions. MemoryGraft~\cite{srivastava2025memorygraft} shows that entries can carry provenance metadata describing where they came from and how they were produced. The Memory Poisoning Attack and Defense framework~\cite{sunil2026memorypoisoning} further shows that authenticated writes and provenance checks can help detect and quarantine compromised entries.

The implication is that memory integrity must be treated as a platform concern rather than a prompt-level defense. In open agentic systems, persistent memory writes should carry provenance linking them to the originating intent, tool invocation, and authorization context, and compromised or unattributable entries should be quarantined rather than silently reused.

\smallskip
\noindent\textbf{Evaluate platforms, not just attack success.}
\label{sec:principles:evaluation}
A repeated weakness in the literature is that evaluation often stops at attack success rate or task completion. These metrics matter, but they say little about whether the platform remains governable once failures occur. Secure agent ecosystems therefore need evaluation criteria for platform properties such as capability scoping, provenance completeness, revocation, auditability, and recovery.

Existing benchmarks are increasingly good at reproducing attacks and comparing defenses, but they rarely measure capability overreach under explicit manifests, re-authorization of plan mutation, provenance preservation across tool and memory flows, or recovery after compromise. Benchmark maturity can therefore overstate platform maturity.

The design principle is to evaluate the surrounding control system, not just the model or task outcome. A secure platform should be assessed not only by whether it blocks an exploit, but also by whether it scopes authority correctly, records attributable state transitions, supports revocation, and remains recoverable after failure.




\begin{figure}[!t]
  \centering
  \includegraphics[width=0.99\linewidth]{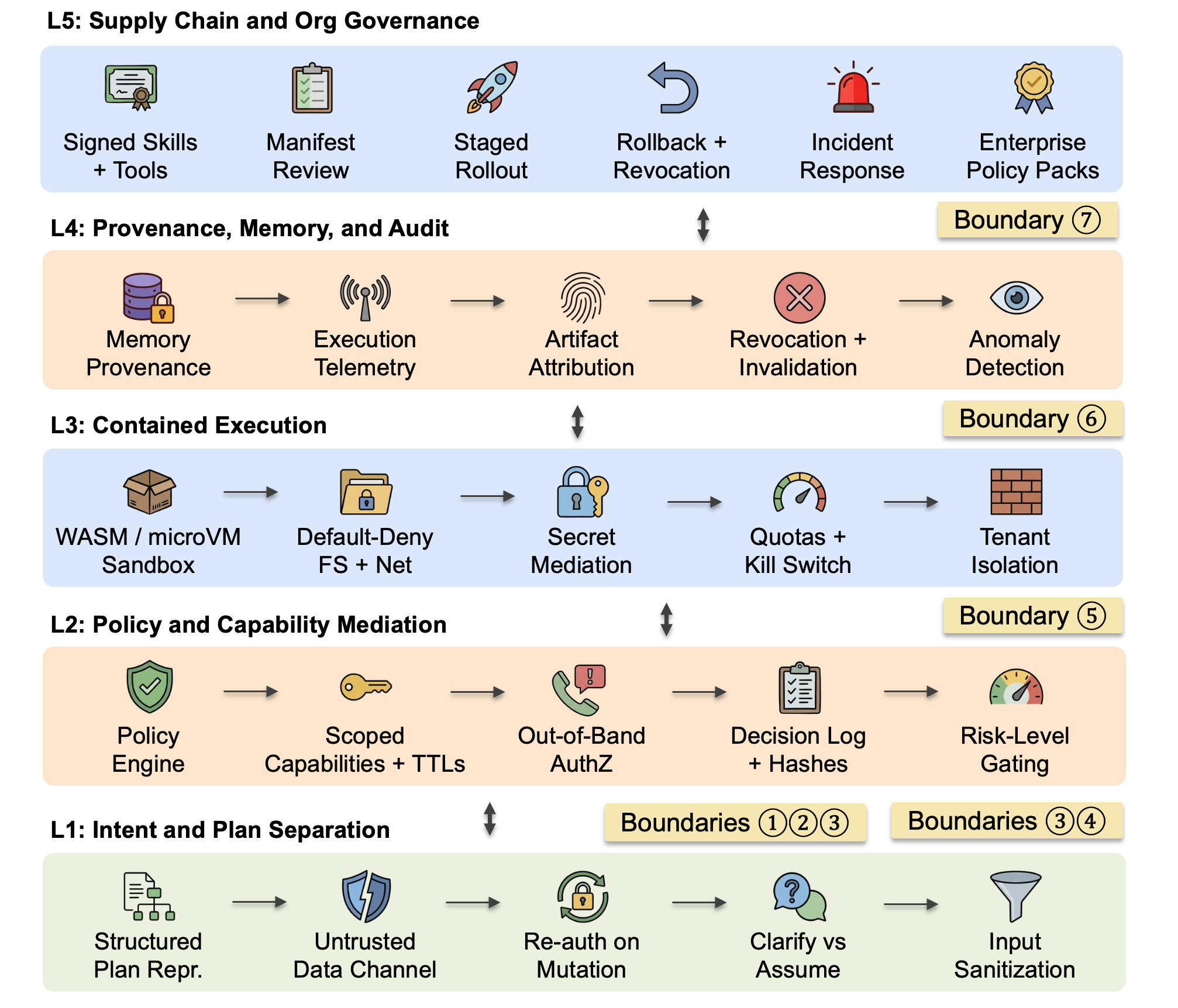}
  \vspace{-0.1in}
  \caption{Systematized reference doctrine}\vspace{-0.1in}
  \label{fig:reference-doctrine}
  \vspace{-0.1in}
\end{figure}

\subsection{Reference Doctrine}
\label{sec:principles:doctrine}

We now compose the distilled principles into a governance scaffold for secure-by-construction open agent systems.

The doctrine has five layers, ordered from execution-near controls to organizational governance. It is a compositional synthesis of patterns already validated in parts of the literature, not a claim that a complete five-layer implementation has already been realized in practice. \Cref{fig:reference-doctrine} summarizes the full doctrine.

\textit{\underline{Layer 1}: intent and plan separation.}
\label{sec:principles:doctrine:l1}
Layer~1 governs the transition from user intent to agent planning, covering the User--Agent, Content--Agent, and Planning--Execution boundaries. User intent is translated into a structured plan rather than left implicit in the raw token stream. Untrusted retrieved content stays on a separate data path and cannot be silently promoted to instructions. Plan mutations that expand scope, change targets, or raise privilege require explicit re-authorization. When intent remains ambiguous, the default is clarification rather than silent assumption. This layer is grounded in CaMeL~\cite{debenedetti2025camel}, Task Shield~\cite{jia2024taskshield}, and Fides~\cite{costa2025fides}.

\textit{\underline{Layer 2}: policy and capability mediation}
\label{sec:principles:doctrine:l2} Layer~2 places a deterministic policy engine between planning and tool execution, covering the Planning--Execution and Agent--Tool boundaries. Every tool call is checked against declarative policy before dispatch. Permissions are scoped to specific resources and bounded by time-to-live constraints. High-risk actions, such as credential access, external network egress, or bulk data operations, require out-of-band authorization from the user or organizational policy. Policy decisions are logged with context hashes so that their basis can later be verified. Progent~\cite{shi2025progent}, Fides~\cite{costa2025fides}, and mobile OS permission models~\cite{felt2012androidpermissions} ground this layer.

\textit{\underline{Layer 3}: contained execution.}
\label{sec:principles:doctrine:l3} Layer~3 addresses the Runtime--Host boundary by ensuring that tool code runs inside isolation domains that bound the blast radius of compromise or misuse. Tools execute in WebAssembly modules, microVMs, containers, or process-level sandboxes under default-deny filesystem and network policies. Secrets are injected through mediation layers rather than exposed directly to tool code, and are revoked on completion. Resource quotas and kill switches further limit runaway behavior.

This layer is grounded in WaVe~\cite{johnson2023wave}, Chrome site isolation~\cite{jia2016chromesandbox}, and the family of Visor/Firecracker-style sandboxing models.

\textit{\underline{Layer 4}: provenance, memory integrity, and audit.}
\label{sec:principles:doctrine:l4} Layer~4 addresses the memory-future sessions boundary by treating persistent memory as an integrity-protected store rather than an append-only context buffer. Every memory write carries provenance metadata recording the originating user intent, plan step, and capability token under which it was authorized. Every tool execution emits structured telemetry linking inputs, outputs, and side effects to specific plan steps. If a memory entry or tool is later found to be compromised, revocation can invalidate downstream artifacts and trigger repair procedures.

This layer combines three foundations: signing, end-to-end provenance, and structured observability. Sigstore~\cite{newman2022sigstore} and in-toto~\cite{torresarias2019intoto} provide artifact signing and provenance chains. AgentOps~\cite{dong2024agentops} provides execution tracing. It is further motivated by AgentPoison~\cite{chen2024agentpoison}, MemoryGraft~\cite{srivastava2025memorygraft}, and the memory poisoning defence~\cite{sunil2026memorypoisoning}.

\textit{\underline{Layer 5}: supply chain and organizational governance.}
\label{sec:principles:doctrine:l5} Layer~5 addresses the supply chain-organisation boundary by extending security beyond the runtime to the processes that approve, deploy, monitor, and retire agent skills. Skills and MCP tool servers are signed with verifiable identities. Capability manifests and declared data-access patterns are reviewed before deployment. Rollout follows staged release and canary analysis, with automated rollback on policy violation. Multi-tenant platforms enforce tenant isolation so that a compromised skill in one tenant cannot access another tenant's data or tools. Enterprise policy packs encode organizational requirements as machine-checkable rules, while incident-response playbooks define escalation, emergency disablement, and post-incident review procedures.

This layer is motivated by the MCP Safety Audit~\cite{radosevich2025mcpaudit}, AgentOps~\cite{dong2024agentops}, Seven Security Challenges~\cite{ko2025sevenchallenges}, and Sigstore~\cite{newman2022sigstore}.

\smallskip\noindent\textbf{The doctrine as a governance scaffold.}
The five layers constitute a governance scaffold for operating under persistent uncertainty, not a correctness proof.
Layers 1 and 2 implement bounded delegation: they prevent the agent from acquiring authority beyond what the current task requires.
Layers 3 and 4 implement controllability and attributability: they limit the blast radius of failures and ensure that decisions can be reconstructed after the fact.
Layer 5 implements organizational recoverability: it defines how the platform responds, escalates, and recovers when compromise occurs.
No deployed system yet implements all five layers simultaneously, and some of the underlying standards (capability manifests, signed tool descriptions, provenance schemas) remain to be specified.
The doctrine therefore marks the boundary between what current research has enabled and what the field still needs to build.

\section{Research Agenda}
\label{sec:agenda}

This section shows an evaluation scorecard for platform security posture and a concise set of open research directions. 

\smallskip
\noindent\textbf{Specification and requirements.}
Current systems still lack a principled way to declare what an agent may do, what it must never do, and when it should stop and ask for clarification. The corpus reflects this gap: capability overreach is not directly measured in existing benchmarks, and explicit permission manifests are largely absent from evaluated systems. Open questions include how to define machine-checkable permissions, how to represent ambiguity thresholds and prohibited side effects, and how to handle controlled privilege escalation when a task legitimately requires it. Without such mechanisms, deployed agents tend to inherit ambient authority by default.

\smallskip
\noindent\textbf{Architecture and platforms.}
The defense coverage matrix (Table~\ref{tab:defense-coverage}) shows that \emph{memory integrity} receives 0-of-12 coverage and \emph{runtime containment} only 2-of-12, while the three operational-maturity columns are almost uniformly empty.
The reference doctrine in \Cref{sec:principles} synthesizes layered isolation from several independent proposals, yet no production system implements all five layers simultaneously.
Key open problems include defining the minimal trusted computing base for an agent platform, determining the right isolation boundaries between planning, memory, and execution subsystems, and designing safe abstractions for multi-agent delegation chains where each agent holds different capability scopes.
Empirical studies comparing sandboxing strategies (containers, WebAssembly, microVMs, brokered execution) under realistic agent workloads remain absent, consistent with the single deployment-experience paper in the entire 50-paper corpus (\S\ref{sec:synthesis:quantitative}).

\smallskip
\noindent\textbf{Testing and verification.}
The 6-of-8 scorecard metrics with zero benchmark coverage (\S\ref{sec:synthesis:quantitative}) and the concentration of all 10 benchmarks on the \textsc{Test} lifecycle phase (\S\ref{sec:synthesis:thematic}) reveal that current evaluation infrastructure is structurally optimized for measuring \emph{whether} agents fail rather than \emph{whether} agent platforms are well-engineered. Existing benchmarks such as BIPIA~\cite{yi2024bipia}, InjecAgent~\cite{zhan2024injecagent}, AgentDojo~\cite{debenedetti2024agentdojo}, and ASB~\cite{zhang2025asb} focus primarily on prompt injection and measure attack success rate against benign-task utility. However, no benchmark currently tests cross-session persistence, supply-chain compromise, or incident recovery.

Open questions include how to construct regression suites that capture adaptive and chained attacks, whether safety invariants over tool-call sequences and memory mutations can be formalized and checked statically, and how fuzz-testing techniques from systems security can be adapted to stochastic planning.

To close these blind spots, we propose seven evaluation dimensions that future benchmarks should address:

\begin{packeditemize}
  \item \emph{Lifecycle completeness.} Benchmarks should evaluate across \textsc{Spec}, \textsc{Design}, \textsc{Deploy}, and \textsc{Ops} phases, not only \textsc{Test}, since the thin population of the \textsc{Deploy} phase (7 of 50 papers) and \textsc{Impl} phase (5 of 50) confirms that pre-release and release-time security remain structurally untested.
  \item \emph{Deployment realism.} Evaluation should occur under production-grade configurations and default permission settings, because no current benchmark assesses out-of-the-box deployment posture.
  \item \emph{Cross-session attacks}: benchmarks must model memory poisoning that persists across sessions, given that all 10 existing benchmarks are single-session by design and \emph{Memory Cross-Session Persistence} receives at most partial coverage from one benchmark in the metrics scorecard from \S\ref{sec:synthesis:quantitative}.
  \item \emph{Supply-chain attacks} (malicious skill or plugin installation through registries). Evaluation scenarios should address the near-total absence of the \textsc{Registry-SupplyChain} trust boundary (2 of 50 papers) from current benchmark suites.
  \item Benchmarks should measure \emph{incident recovery}, including Mean Time to Recovery and forensic attributability; both \emph{MTTR} and \emph{Operational Audit Breadth} receive 0-of-10 benchmark coverage.
  \item \emph{Organizational governance scenarios}, including multi-tenant, role-based access, and approval-workflow settings, should be benchmarked to reflect the fact that the \textsc{Org-Operator} trust boundary appears in only 3 of 50 papers.
  \item \emph{Mixed AI-plus-software attacks.} Compound attack chains that cross model and infrastructure layers (\eg prompt injection combined with parser vulnerabilities and overprivileged tool configurations) should be evaluated, because no existing benchmark tests attacks that span more than one trust boundary simultaneously.
  
\end{packeditemize}

\smallskip
\noindent\textbf{Supply chain and governance.}
The \textsc{Registry-SupplyChain} trust boundary appears in only 2 of 50 papers (both adjacent engineering foundations rather than agent-security studies), and \emph{supply-chain provenance} is addressed by only 1 of 12 defense systems in Table~\ref{tab:defense-coverage}.
This near-total gap confirms that the rapid growth of open skill registries introduces software supply-chain risks that the agent-security literature has barely begun to address.
Critical questions include what governance model skill registries should adopt (centralized review, community attestation, or hybrid), how organizations should revoke or quarantine compromised tools without disrupting dependent workflows, and how open-ecosystem incentives can be aligned with enterprise policy requirements such as audit trails and tenant isolation.
Lessons from package-manager security in conventional software ecosystems~\cite{newman2022sigstore} offer a starting point, but the executable-description attack surface unique to agent tools demands new mitigations.

\smallskip
\noindent\textbf{Human factors and operations.}
The corpus contains only 2 operations papers and 3 audits among 50 studies (\S\ref{sec:synthesis:quantitative}), and deployment experience is represented by a single paper (the Gemini defense retrospective).
The \textsc{Org-Operator} trust boundary, where human oversight meets agent autonomy, appears in only 3 of 50 papers, making it the second-least-studied boundary after \textsc{Registry-SupplyChain}.
Approval workflows represent a structural tension: too much friction reduces adoption, while too little friction permits unsafe actions.
Research is needed on how to calibrate approval granularity so that routine operations proceed without interruption while high-risk actions trigger out-of-band confirmation.
Equally important are observability primitives that enable operators to investigate incidents involving partially autonomous systems, and post-compromise playbooks that specify how to attribute, contain, and remediate agent-mediated breaches.

\smallskip
\noindent\textbf{Standardization opportunities.}
The fragmentation visible across the corpus points toward standardization as a precondition for cumulative progress: 37 of 50 papers remain preprints, evaluation metrics are inconsistent, and no two defense systems share a common policy language (\S\ref{sec:synthesis:defense-coverage}).
Concrete candidates include a capability-manifest schema that declares tool permissions, data scopes, and network access in a portable format; a signed tool-description format that binds executable behavior to a verifiable identity; a provenance schema for agent execution traces that records planner decisions, capability tokens, and tool outputs; a security-profile specification for MCP-like protocol ecosystems; and a reporting standard for agent-security benchmarks that requires lifecycle coverage, deployment realism, and cross-session attack metrics alongside conventional ASR and utility scores.

\section{Discussions}
\label{sec-discussion}

We close with two discussion points. The first demonstrates the utility of the taxonomy through a concrete incident, and the second examines the main threats to validity and the scope limitations.

\subsection{Taxonomy Utility Demonstration}
\label{sec:synthesis:casestudy}

To illustrate the practical utility of the taxonomy, we revisit the credential-exfiltration scenario introduced earlier ~\cite{yi2026tamingopenclaw, debenedetti2025camel, radosevich2025mcpaudit, fu2024imprompter} and compare two ways of analyzing it: a conventional threat-centric reading and a lifecycle-aware software-engineering reading.

\begin{center}
   \colorbox{orange!11}{
    \begin{minipage}{0.92\linewidth}
(\textbf{Scenario}) A developer uses a coding agent with \mcp tool access to summarize recent GitHub issues and update project documentation. One issue contains adversarial instructions hidden in a code comment. The agent, running with the developer's ambient filesystem and network permissions, interprets the malicious content, reads local credentials from \texttt{\~{}/.aws/credentials}, and exfiltrates them through an HTTP tool call. The compromise remains unnoticed for days because the logs record only seemingly normal tool invocations.
   \end{minipage}
}
\end{center}

A conventional attack taxonomy would classify the incident as \emph{indirect prompt injection} at the \textsc{Content-Agent} boundary. This diagnosis is correct but incomplete. It identifies the immediate trigger, but not the architectural conditions that allowed the attack to progress from malicious content to credential theft. The natural mitigations would therefore focus on prompt hardening, input filtering, or adversarial fine-tuning.

\begin{figure}[t]
  \centering
  \includegraphics[width=0.98\columnwidth]{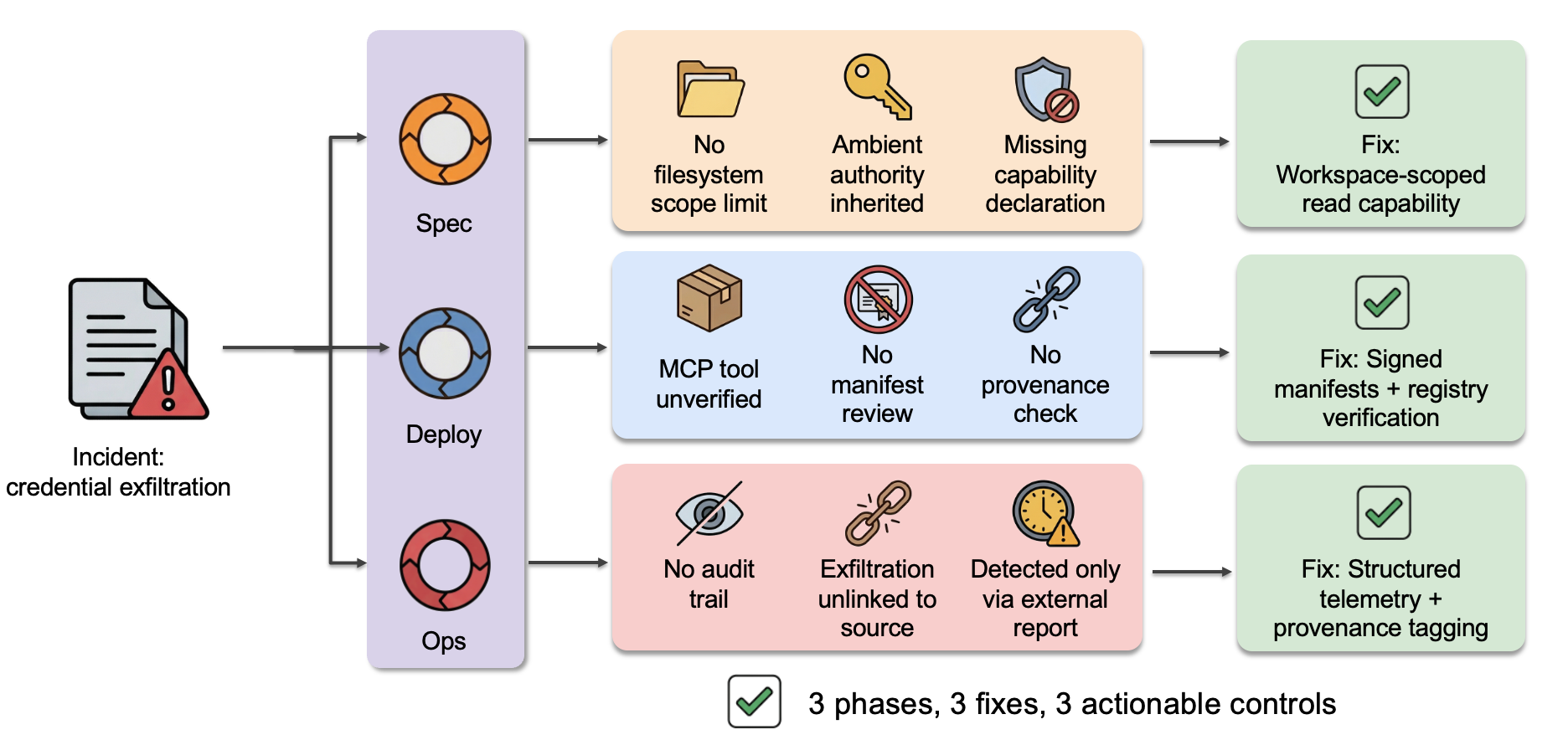}
  \vspace{-0.1in}
  \caption{Case study: the same credential-exfiltration incident analyzed under our taxonomy}
  \vspace{-0.1in}
  \label{fig:case-study}
\end{figure}

In comparison, our taxonomy instead reveals three contributing failures across different lifecycle phases (Fig.~\ref{fig:case-study}):

\begin{packeditemize}
  \item \textit{Specification failure (\textsc{Spec}).}
  No capability declaration restricted filesystem access to the project workspace. The agent inherited ambient authority and could read sensitive files outside task scope. A correct specification would limit reads to the workspace and deny access to credential paths.

  \item \textit{Deployment failure (\textsc{Deploy}).}
  The \mcp tool server was installed without provenance checks or manifest review. No mechanism verified its origin, integrity, or declared capabilities. A stronger deployment policy would require signed manifests and registry-level provenance validation.

  \item \textit{Operations failure (\textsc{Ops}).}
  The system logged tool invocations but did not preserve the provenance needed to connect the credential read and the outbound HTTP request. As a result, the attack chain was hard to reconstruct. An operational control would emit structured telemetry with provenance annotations to support attribution and detection.
\end{packeditemize}

We ask whether capabilities were properly scoped, whether the tool supply chain was verified, and whether the incident could be detected and attributed. These questions map directly to the lifecycle phases and trust boundaries, yielding concrete engineering requirements rather than model-level robustness goals.

\subsection{Threats to Validity}
\label{subsec:threats}

We also consider threats and limitations to validity. 

\textit{Construct validity.}
The boundary between autonomous agents and tool-augmented LLM applications is not always clean. Some systems differ from retrieval-augmented generation mainly by adding a planning loop, whereas others include persistent memory, multi-step tool use, and inter-agent delegation. We apply the inclusion criteria in \Cref{sec:methodology:corpus} and code each paper along standardized dimensions, but borderline cases remain. Benchmark success rates also do not map directly to deployment risk, since controlled evaluations rarely capture the full complexity of production environments.

\textit{Internal validity.}
Coding across seven dimensions inevitably involves judgment. This is especially true when a paper spans multiple lifecycle phases or control loci, or when the original work uses terminology that does not align neatly with the survey vocabulary. Iterative normalization reduces some of this ambiguity, but single-analyst coding still limits inter-rater reliability. The absence of a second independent coder means formal inter-rater agreement (such as Cohen's~$\kappa$) cannot be reported; future replications should include at least one independent coder with kappa measurement on a random 20\% sample.

\textit{External validity.}
The findings are grounded mainly in open agentic systems with open tool registries, protocol-mediated integrations, and cloud-hosted LLM backends. Browser agents, on-device mobile agents, and embodied or robotic agents operate under different resource constraints and trust assumptions, so some conclusions may need adaptation before they transfer cleanly. The underlying platforms are also evolving quickly, which means some attack surfaces and mitigations may shift as architectures mature.

\textit{Scope limitation.}
Several adjacent areas are excluded. Training-time model poisoning and data poisoning are out of scope because they primarily target the model provider rather than the agent platform. Pure alignment or jailbreak studies without tool use are excluded because they do not exercise the trust boundaries central to this paper. Hardware side channels and broader AI ethics questions are likewise outside the software-engineering focus.

\section{Conclusion}
\label{sec:conclusion}

We assess the security of open agentic systems through a software engineering lens. The synthesis reveals three persistent blind spots: supply-chain integrity, persistent memory, and operational governance. We distil design principles, a  reference doctrine, and a research agenda for secure-by-construction open agent systems.

\begin{table*}[!t]
\centering
\caption{Cross-Paper Synthesis. The table combines venue metadata with eight dimensions: publication status, type, agent setting, primary concern, trust boundary, lifecycle phase, practical artifact, and evidence type. Control locus is analyzed separately in Table~\ref{tab:defense-coverage}, since many papers span multiple loci.}
\label{tab:synthesis}
\scriptsize
\setlength{\tabcolsep}{3pt}
\begin{threeparttable}
\renewcommand{\arraystretch}{1.0}
\begin{tabular}{@{}c | r c c c| c l | l l | l c@{}}
\toprule
\rowcolor{orange!20}
\multicolumn{1}{c}{\textbf{Group}} & 
\multicolumn{1}{c}{\textbf{Paper}} & 
\multicolumn{1}{c}{\textbf{Year/Venue}} & 
\multicolumn{1}{c}{\textbf{Status}} & 
\multicolumn{1}{c}{\textbf{Type}} & 
\multicolumn{1}{c}{\textbf{Agent}} & 
\multicolumn{1}{c}{\textbf{Concern}} & 
\multicolumn{1}{c}{\textbf{Trust Boundary}} & 
\multicolumn{1}{c}{\textbf{Lifecycle}} & 
\multicolumn{1}{c}{\textbf{Artifact}} & 
\multicolumn{1}{c}{\textbf{Evid.}} \\
\midrule
\multirow{4}{*}{\makecell[c]{Survey\\ \& Broad\\ Threat\\ Overview}}
& Taming \openclaw~\cite{yi2026tamingopenclaw} & 2026/arXiv & Pre & Survey & OC & ecosystem threats & Ecosystem-Wide & Design+Deploy+Ops & ecosystem synth. & Sv \\
& Navigating Risks~\cite{gan2024navigatingrisks} & 2024/arXiv & Pre & Survey & GA & sec./privacy risks & Ecosystem-Wide & Design+Ops & survey taxonomy & Sv \\
& Trustworthy Agents~\cite{yu2025trustworthyagents} & 2025/arXiv & Pre & Survey & GA & trust modules & Ecosystem-Wide & Design+Ops & modular taxonomy & Sv \\
& Securing Agentic AI~\cite{narajala2025securingagenticai} & 2025/arXiv & Pre & ThreatModel & EA & enterprise threats & Ecosystem-Wide & Design+Deploy+Ops & threat model & Ps \\
\midrule

\multirow{10}{*}{\makecell[c]{Benchmark\\ \& Empirical\\ Evaluation}}
& BIPIA~\cite{yi2024bipia} & 2024/MSR & Lab & Benchmark & TA & indirect prompt inj. & Content-Agent & Test & bench suite & Bm \\
& InjecAgent~\cite{zhan2024injecagent} & 2024/ACL & P & Benchmark & TA & tool-aided PI & Content-Agent & Test & 1054-case bench & Bm \\
& AgentDojo~\cite{debenedetti2024agentdojo} & 2024/NeurIPS & P & Benchmark & TA & dyn. PI + defenses & Content-Agent & Test & dynamic env. & Bm \\
& ASB~\cite{zhang2025asb} & 2025/ICLR & P & Benchmark & GA & multi-stage attacks & Ecosystem-Wide & Test & bench + metrics & Bm \\
& ToolEmu~\cite{ruan2024toolemu} & 2024/ICLR & P & Benchmark & TA & tool-use risks & Agent-Tool & Test & emulated sandbox & Bm \\
& WASP~\cite{evtimov2025wasp} & 2025/arXiv & Pre & Benchmark & MmA & web-agent PI & Content-Agent & Test+Exec & web benchmark & Bm \\
& AgentDAM~\cite{zharmagambetov2025agentdam} & 2025/arXiv & Pre & Benchmark & MmA & web privacy leakage & Content-Agent & Test+Exec & privacy benchmark & Bm \\
& WAInjectBench~\cite{liu2025wainjectbench} & 2025/arXiv & Pre & Benchmark & MmA & web PI detection & Content-Agent & Test & detect. benchmark & Bm \\
& SimPrivacy~\cite{zhang2025simprivacy} & 2025/arXiv & Pre & Benchmark & GA & privacy-risk discovery & Content-Agent & Test & simulation bench & Bm \\
& AGENTFUZZER~\cite{wang2025agentfuzzer} & 2025/arXiv & Pre & Attack & TA & black-box PI fuzzing & Content-Agent & Test & fuzzing framework & ES \\
\midrule

\multirow{13}{*}{\makecell[c]{Defense\\ \& Architecture}}
& CaMeL~\cite{debenedetti2025camel} & 2025/arXiv & Pre & Defense & TA & control/data confusion & Content-Agent & Design+Exec & capability arch. & Pt \\
& Fides~\cite{costa2025fides} & 2025/arXiv & Pre & Defense & TA & conf./int. in tool use & Agent-Tool & Design+Exec & formal model + sys. & FM \\
& Progent~\cite{shi2025progent} & 2025/arXiv & Pre & Defense & TA & overprivilege & Agent-Tool & Design+Exec & privilege controller & Pt \\
& Task Shield~\cite{jia2024taskshield} & 2024/arXiv & Pre & Defense & TA & action drift & Content-Agent & Design+Exec & alignment defense & Pt \\
& GuardAgent~\cite{xiang2024guardagent} & 2025/proj.\ page & Lab & Defense & TA & executable guardrails & Agent-Tool & Design+Exec & guardrail agent & Pt \\
& Imprompter~\cite{fu2024imprompter} & 2024/arXiv & Pre & Attack & TA & improper tool use & Agent-Tool & Exec & attack suite & ES \\
& Gemini Lessons~\cite{shi2025geminidefense} & 2025/Google & Lab & Ops & EA & prod. defense loops & Content-Agent & Test+Ops & prod. retro. & DE \\
& AgentSpec~\cite{wang2025agentspec} & 2025/arXiv & Pre & Defense & TA & runtime enforcement & Agent-Tool & Impl+Exec & enforcement frmwk & Pt \\
& Pro2Guard~\cite{wang2025pro2guard} & 2025/arXiv & Pre & Defense & TA & proactive runtime safety & Agent-Tool & Impl+Exec & checked enforcement & FM \\
& RTBAS~\cite{zhong2025rtbas} & 2025/arXiv & Pre & Defense & TA & PI + privacy leakage & Content-Agent & Design+Exec & defense frmwk & Pt \\
& AutoSafeCoder~\cite{nunez2024autosafecoder} & 2024/arXiv & Pre & Defense & EA & secure code generation & Agent-Tool & Impl+Test & secure coding workflow & Pt \\
& PrivWeb~\cite{zhang2025privweb} & 2025/arXiv & Pre & Defense & MmA & web-agent privacy & Content-Agent & Design+Exec & privacy mediation & Pt \\
& LLM Security~\cite{zhang2025securityprinciples} & 2025/arXiv & Pre & Perspective & GA & secure-design principles & Ecosystem-Wide & Design+Ops & principle framing & Ps \\
\midrule

\multirow{3}{*}{\makecell[c]{Memory\\ \& Persistent\\ State Security}}
& AgentPoison~\cite{chen2024agentpoison} & 2024/arXiv & Pre & Attack & MA & poisoned memory & Memory-Future & Memory+Test & attack evaluation & ES \\
& MemoryGraft~\cite{srivastava2025memorygraft} & 2025/arXiv & Pre & Attack & MA & persistent compromise & Memory-Future & Memory+Ops & attack demo & ES \\
& Mem.\ Poison Atk/Def~\cite{sunil2026memorypoisoning} & 2026/arXiv & Pre & Defense & MA & memory poisoning & Memory-Future & Memory+Test & attack/defense study & Pt \\
\midrule

\multirow{5}{*}{\makecell[c]{Multi-Agent\\ \& Coordination\\ Security}}
& MAS Malicious Code~\cite{triedman2025masmaliciouscode} & 2025/arXiv & Pre & Attack & MuA & malicious orchestration & Agent-Agent & Exec+Ops & attack study & ES \\
& Communication Atks~\cite{he2025communicationattacks} & 2025/arXiv & Pre & Attack & MuA & comm.-channel attacks & Agent-Agent & Test+Ops & red-team study & ES \\
& IP Leakage~\cite{wang2025ipleakage} & 2025/arXiv & Pre & Attack & MuA & cross-agent leakage & Agent-Agent & Exec+Ops & leakage study & ES \\
& 7 Security Challenges~\cite{ko2025sevenchallenges} & 2025/arXiv & Pre & Perspective & MuA & multi-agent challenges & Org-Operator & Design+Ops & challenge framing & Ps \\
& Agent Smith~\cite{gu2024agentsmith} & 2024/arXiv & Pre & Attack & MmA & multimodal jailbreak & Content-Agent & Input+Exec & attack study & ES \\
\midrule

\multirow{10}{*}{\makecell[c]{Ecosystem,\\ Protocol, \&\\ Operations}}
& MCP Safety Audit~\cite{radosevich2025mcpaudit} & 2025/arXiv & Pre & Audit & PE & protocol exploits & Protocol-Tool & Impl+Deploy & ecosystem audit & ES \\
& MCP Landscape~\cite{hou2025mcplandscape} & 2025/arXiv & Pre & Survey & PE & MCP threats & Protocol-Tool & Design+Ops & protocol survey & Sv \\
& MCP-Guard~\cite{xing2025mcpguard} & 2025/arXiv & Pre & Defense & PE & protocol integrity & Protocol-Tool & Impl+Exec & defense frmwk & Pt \\
& MCP Red Teaming~\cite{he2025mcpredteam} & 2025/arXiv & Pre & Benchmark & PE & MCP red teaming & Protocol-Tool & Test+Ops & red-team workflow & Bm \\
& AgentOps~\cite{dong2024agentops} & 2024/arXiv & Pre & Ops & GA & observability & Org-Operator & Ops & observability tooling & Pt \\
& EIA~\cite{liao2025eia} & 2025/ICLR & P & Attack & MmA & env. inj. + privacy & Content-Agent & Input+Exec & web attack study & ES \\
& EnvInjection~\cite{wang2025envinjection} & 2025/arXiv & Pre & Attack & MmA & env. prompt inj. & Content-Agent & Input+Exec & web attack study & ES \\
& PrivAgent~\cite{nie2024privagent} & 2024/arXiv & Pre & Attack & GA & privacy leakage & Content-Agent & Test+Ops & red-team workflow & ES \\
& AgentAuditor~\cite{luo2025agentauditor} & 2025/arXiv & Pre & Audit & GA & agent auditing & Ecosystem-Wide & Test+Ops & eval protocol & ES \\
& Disclosure Audits~\cite{das2025disclosureaudits} & 2025/arXiv & Pre & Audit & GA & disclosure audit & Org-Operator & Test+Ops & audit framework & ES \\
\midrule

\multirow{5}{*}{\makecell[c]{Adjacent\\ Engineering\\ Foundations}}
& Sigstore~\cite{newman2022sigstore} & 2022/CCS & P & Foundation & AF & signing + provenance & Registry-SC & Deploy+Ops & signing infra. & Fn \\
& in-toto~\cite{torresarias2019intoto} & 2019/USENIX Sec & P & Foundation & AF & supply-chain integrity & Registry-SC & Deploy+Ops & provenance frmwk & Fn \\
& WaVe~\cite{johnson2023wave} & 2023/IEEE S\&P & P & Foundation & AF & WASM sandboxing & Runtime-Host & Design+Exec & verified sandbox & Fn \\
& Chrome Sandbox~\cite{jia2016chromesandbox} & 2016/CCS & P & Foundation & AF & sandbox boundary & Runtime-Host & Design+Deploy & sandbox study & ES \\
& VRASED~\cite{nunes2019vrased} & 2019/USENIX Sec & P & Foundation & AF & remote attestation & Runtime-Host & Deploy+Ops & verified attestation & FM \\
\bottomrule
\end{tabular}

\begin{tablenotes}
\item Publication status: \textbf{P}~=~Published, \textbf{Lab}~=~Lab publication or workshop paper, \textbf{Pre}~=~Preprint.
\item Agent settings: \textbf{OC}~=~OpenClawEcosystem, \textbf{GA}~=~GeneralAgent, \textbf{TA}~=~ToolAgent, \textbf{EA}~=~EnterpriseAgent, \textbf{MA}~=~MemoryAgent, \textbf{MuA}~=~MultiAgent, \textbf{PE}~=~ProtocolEcosystem, \textbf{MmA}~=~MultimodalAgent, \textbf{AF}~=~AdjacentFoundation.
\item Trust-boundary labels use the canonical vocabulary; \textbf{Registry-SC} denotes Registry-SupplyChain.
\item Evidence types: \textbf{Sv}~=~Survey, \textbf{Bm}~=~Benchmark, \textbf{Pt}~=~Prototype, \textbf{ES}~=~EmpiricalStudy, \textbf{DE}~=~DeploymentExperience, \textbf{FM}~=~FormalModel, \textbf{Ps}~=~Perspective, \textbf{Fn}~=~Foundation.
\end{tablenotes}
\end{threeparttable}
\end{table*}
\begin{table*}[!t]
\centering
\caption{Companion Analytics: SE Lessons and Residual Gaps.  }
\label{tab:synthesis-gaps}
\scriptsize
\setlength{\tabcolsep}{3pt}
\renewcommand{\arraystretch}{1.08}
\begin{tabular}{@{}c | r  r l @{}}
\toprule
\rowcolor{orange!20}
\multicolumn{1}{c}{\textbf{Group}}
& \multicolumn{1}{c}{\textbf{Paper}} 
& \multicolumn{1}{c}{\textbf{SE Lesson}} 
& \multicolumn{1}{c}{\textbf{Residual Gap}} \\
\midrule

\multirow{4}{*}{\makecell[c]{Survey\\ \& Broad\\ Threat\\ Overview}}
& Taming \openclaw~\cite{yi2026tamingopenclaw}
  & Ecosystem studies benefit from lifecycle-oriented synthesis
  & Narrower than a cross-paper SE SoK \\
& Navigating Risks~\cite{gan2024navigatingrisks}
  & Broad surveys need sharper engineering decomposition
  & Limited operational and deployment detail \\
& Trustworthy Agents~\cite{yu2025trustworthyagents}
  & Module taxonomies should be complemented with lifecycle taxonomies
  & Weak release engineering and ops framing \\
& Securing Agentic AI~\cite{narajala2025securingagenticai}
  & Threat models need lifecycle mapping and empirical triangulation
  & Limited evidence synthesis across lifecycle phases \\
\midrule

\multirow{10}{*}{\makecell[c]{Benchmark\\ \& Empirical\\ Evaluation}}
& BIPIA~\cite{yi2024bipia}
  & Early benchmarks shaped attention toward content-boundary attacks
  & Weak deployment realism and ops coverage \\
& InjecAgent~\cite{zhan2024injecagent}
  & Tool-using agents need dedicated evaluation beyond base LLMs
  & Weak on ops, governance, and provenance \\
& AgentDojo~\cite{debenedetti2024agentdojo}
  & Common benchmark infrastructure accelerates defense research
  & Not a deployment doctrine or lifecycle guide \\
& ASB~\cite{zhang2025asb}
  & Evaluation is broadening but still not lifecycle-complete
  & Limited release-engineering and governance analysis \\
& ToolEmu~\cite{ruan2024toolemu}
  & Scalable risk discovery complements direct exploit benchmarks
  & Weak on deployment governance and post-compromise ops \\
& WASP~\cite{evtimov2025wasp}
  & Browser-agent security needs dedicated end-to-end evaluation
  & Web-focused; limited beyond prompt injection \\
& AgentDAM~\cite{zharmagambetov2025agentdam}
  & Web-agent security needs privacy-aware evaluation beyond task success
  & Narrower privacy focus; no lifecycle treatment \\
& WAInjectBench~\cite{liu2025wainjectbench}
  & Browser-agent detection deserves dedicated benchmarking
  & Focused on detection rather than full lifecycle \\
& SimPrivacy~\cite{zhang2025simprivacy}
  & Simulation can scale privacy-risk discovery before live deployment
  & Limited deployment realism \\
& AGENTFUZZER~\cite{wang2025agentfuzzer}
  & Fuzzing should become part of secure-agent regression practice
  & Mainly content-boundary focused \\
\midrule

\multirow{13}{*}{\makecell[c]{Defense\\ \& Architecture}}
& CaMeL~\cite{debenedetti2025camel}
  & Instruction/data separation is a first-class architectural pattern
  & One mechanism family; no ecosystem composition \\
& Fides~\cite{costa2025fides}
  & Deterministic policy mediation is plausible and rigorous
  & Limited supply-chain and registry treatment \\
& Progent~\cite{shi2025progent}
  & Least privilege can be made programmable and dynamic
  & Limited ops and post-deployment framing \\
& Task Shield~\cite{jia2024taskshield}
  & Task alignment is a usable action-filtering invariant
  & Narrow attack model and limited deployment framing \\
& GuardAgent~\cite{xiang2024guardagent}
  & Policy translation into executable checks is viable
  & Limited deployment and ecosystem-level framing \\
& Imprompter~\cite{fu2024imprompter}
  & Tool misuse deserves its own failure class
  & Attack-centric; no release or ops guidance \\
& Gemini Lessons~\cite{shi2025geminidefense}
  & Production defense requires continuous evaluation loops
  & Product-specific; not generalized to a doctrine \\
& AgentSpec~\cite{wang2025agentspec}
  & Runtime enforcement is becoming programmable and reusable
  & Limited ecosystem-level treatment \\
& Pro2Guard~\cite{wang2025pro2guard}
  & Formal runtime enforcement can complement policy mediation
  & Early-stage; preprint-only evidence \\
& RTBAS~\cite{zhong2025rtbas}
  & Defense stacks increasingly combine prompt and tool mediation
  & Still scoped to guardrail logic \\
& AutoSafeCoder~\cite{nunez2024autosafecoder}
  & Code-agent security can inherit static analysis and fuzzing
  & Coding-agent specific \\
& PrivWeb~\cite{zhang2025privweb}
  & Browser-agent defenses need privacy-aware mediation layers
  & Web-focused; no cross-ecosystem treatment \\
& LLM Security Principles~\cite{zhang2025securityprinciples}
  & Security-principle framing aligns with SE doctrine
  & Less systematic than a SoK \\
\midrule

\multirow{3}{*}{\makecell[c]{Memory\\ \& Persistent\\ State Security}}
& AgentPoison~\cite{chen2024agentpoison}
  & Memory is mutable security-critical state
  & Weak lifecycle translation; no platform controls \\
& MemoryGraft~\cite{srivastava2025memorygraft}
  & Persistent compromise is an ecosystem-state problem
  & Defenses and provenance design still immature \\
& Mem.\ Poison Atk/Def~\cite{sunil2026memorypoisoning}
  & Memory security needs realistic domain evaluation
  & Domain-specific; not generalized to ecosystem doctrine \\
\midrule

\multirow{5}{*}{\makecell[c]{Multi-Agent\\ \& Coordination\\ Security}}
& MAS Malicious Code~\cite{triedman2025masmaliciouscode}
  & Delegation amplifies risk and needs policy controls
  & Limited design guidance for secure delegation \\
& Communication Atks~\cite{he2025communicationattacks}
  & Communication channels need security semantics
  & Narrow attack family; no contract or provenance \\
& IP Leakage~\cite{wang2025ipleakage}
  & Information-flow remains central in multi-agent settings
  & Confidentiality-focused; weak on broader lifecycle \\
& 7 Security Challenges~\cite{ko2025sevenchallenges}
  & Cross-domain agents require governance-aware reasoning
  & Lacks corpus synthesis and empirical validation \\
& Agent Smith~\cite{gu2024agentsmith}
  & Non-text inputs belong in the security taxonomy
  & Multimodal-specific; limited generalization \\
\midrule

\multirow{10}{*}{\makecell[c]{Ecosystem,\\ Protocol, \&\\ Operations}}
& MCP Safety Audit~\cite{radosevich2025mcpaudit}
  & Tool protocols create a new supply-chain surface
  & MCP-specific; not a lifecycle synthesis \\
& MCP Landscape~\cite{hou2025mcplandscape}
  & Standards work is becoming central to ecosystem security
  & Specific to MCP; limited cross-protocol view \\
& MCP-Guard~\cite{xing2025mcpguard}
  & Tool protocols need explicit integrity mediation
  & Mechanism-focused; no ecosystem governance \\
& MCP Red Teaming~\cite{he2025mcpredteam}
  & Red teaming should be part of release engineering
  & Narrow to testing phase \\
& AgentOps~\cite{dong2024agentops}
  & Observability is prerequisite infrastructure for governance
  & Not security-first; no threat model \\
& EIA~\cite{liao2025eia}
  & Adversarial webpages create a distinctive browser-agent surface
  & Focused on browser privacy scenarios \\
& EnvInjection~\cite{wang2025envinjection}
  & Browser environments are active attack surfaces, not passive inputs
  & Narrow to web-agent settings \\
& PrivAgent~\cite{nie2024privagent}
  & Privacy leakage can be operationalized as agent red teaming
  & Privacy-focused; limited lifecycle analysis \\
& AgentAuditor~\cite{luo2025agentauditor}
  & Safety auditing is becoming a reusable engineering primitive
  & Preprint and evaluation-centric \\
& Disclosure Audits~\cite{das2025disclosureaudits}
  & Auditability should itself be benchmarked as a platform capability
  & Privacy-centric; limited deployment treatment \\
\midrule

\multirow{5}{*}{\makecell[c]{Adjacent\\ Engineering\\ Foundations}}
& Sigstore~\cite{newman2022sigstore}
  & Skill/tool ecosystems need provenance and revocation
  & Not agent-specific; requires translation \\
& in-toto~\cite{torresarias2019intoto}
  & Secure ecosystems need pipeline integrity, not only runtime controls
  & Focused on build pipelines, not live orchestration \\
& WaVe~\cite{johnson2023wave}
  & Runtime isolation claims should inherit from real sandboxing work
  & Not agent-specific; needs tool-call adaptation \\
& Chrome Sandbox~\cite{jia2016chromesandbox}
  & Browser isolation is a useful precedent for agent isolation
  & Browser-specific and dated \\
& VRASED~\cite{nunes2019vrased}
  & Provenance and attestation should back audit claims
  & Attestation alone is insufficient for governance \\
\bottomrule
\end{tabular}
\end{table*}

\bibliographystyle{unsrt}
\bibliography{refs}

@misc{yi2026tamingopenclaw,
      title={Taming OpenClaw: Security Analysis and Mitigation of Autonomous LLM Agent Threats}, 
      author={Xinhao Deng and Yixiang Zhang and Jiaqing Wu and Jiaqi Bai and Sibo Yi and Zhuoheng Zou and Yue Xiao and Rennai Qiu and Jianan Ma and Jialuo Chen and Xiaohu Du and Xiaofang Yang and Shiwen Cui and Changhua Meng and Weiqiang Wang and Jiaxing Song and Ke Xu and Qi Li},
      year={2026},
      eprint={2603.11619},
      archivePrefix={arXiv},
      primaryClass={cs.CR},
      url={https://arxiv.org/abs/2603.11619}, 
}

@article{gan2024navigatingrisks,
  title   = {Navigating the Risks: A Survey of Security, Privacy, and Ethics Threats in {LLM}-Based Agents},
  author  = {Gan, Yuyou and Yang, Yong and Ma, Zhe and He, Ping and Zeng, Rui and Wang, Yiming and Li, Qingming and Zhou, Chunyi and Li, Songze and Wang, Ting and Gao, Yunjun and Wu, Yingcai and Ji, Shouling},
  journal = {arXiv preprint arXiv:2411.09523},
  year    = {2024}
}

@inproceedings{yu2025trustworthyagents,
  title     = {A Survey on Trustworthy {LLM} Agents: Threats and Countermeasures},
  author    = {Yu, Miao and Meng, Fanci and Zhou, Xinyun and Wang, Shilong and Mao, Junyuan and Pang, Linsey and Chen, Tianlong and Wang, Kun and Li, Xinfeng and Zhang, Yongfeng and An, Bo and Wen, Qingsong},
  booktitle = {Proceedings of the 31st {ACM} {SIGKDD} Conference on Knowledge Discovery and Data Mining},
  year      = {2025},
  doi       = {10.1145/3711896.3736561}
}

@article{narajala2025securingagenticai,
  title   = {Securing Agentic {AI}: A Comprehensive Threat Model and Mitigation Framework for Generative {AI} Agents},
  author  = {Narajala, Vineeth Sai and Narayan, Om},
  journal = {arXiv preprint arXiv:2504.19956},
  year    = {2025}
}

@inproceedings{yi2024bipia,
  title     = {Benchmarking and Defending Against Indirect Prompt Injection Attacks on Large Language Models},
  author    = {Yi, Jingwei and Xie, Yueqi and Zhu, Bin and K{\i}c{\i}man, Emre and Sun, Guangzhong and Xie, Xing and Wu, Fangzhao},
  booktitle = {Proceedings of the 31st {ACM} {SIGKDD} Conference on Knowledge Discovery and Data Mining},
  year      = {2025},
  doi       = {10.1145/3690624.3709179}
}

@inproceedings{zhan2024injecagent,
  title     = {{InjecAgent}: Benchmarking Indirect Prompt Injections in Tool-Integrated Large Language Model Agents},
  author    = {Zhan, Qiusi and Liang, Zhixiang and Ying, Zifan and Kang, Daniel},
  booktitle = {Findings of the Association for Computational Linguistics: ACL 2024},
  pages     = {10471--10506},
  year      = {2024},
  doi       = {10.18653/v1/2024.findings-acl.624},
  url       = {https://aclanthology.org/2024.findings-acl.624/}
}

@inproceedings{debenedetti2024agentdojo,
author = {Debenedetti, Edoardo and Zhang, Jie and Balunovic, Mislav and Beurer-Kellner, Luca and Fischer, Marc and Tram\`{e}r, Florian},
title = {AgentDojo: a dynamic environment to evaluate prompt injection attacks and defenses for LLM agents},
year = {2024},
isbn = {9798331314385},
publisher = {Curran Associates Inc.},
address = {Red Hook, NY, USA},
booktitle = {Proceedings of the 38th International Conference on Neural Information Processing Systems},
articleno = {2636},
numpages = {26},
location = {Vancouver, BC, Canada},
series = {NIPS '24}
}

@inproceedings{zhang2025asb,
  title     = {Agent Security Bench ({ASB}): Formalizing and Benchmarking Attacks and Defenses in {LLM}-based Agents},
  author    = {Zhang, Hanrong and Huang, Jingyuan and Mei, Kai and Yao, Yifei and Wang, Zhenting and Zhan, Chenlu and Wang, Hongwei and Zhang, Yongfeng},
  booktitle = {International Conference on Learning Representations ({ICLR})},
  year      = {2025},
  url       = {https://openreview.net/forum?id=V4y0CpX4hK}
}

@inproceedings{ruan2024toolemu,
  title     = {Identifying the Risks of {LM} Agents with an {LM}-Emulated Sandbox},
  author    = {Ruan, Yangjun and Dong, Honghua and Wang, Andrew and Pitis, Silviu and Zhou, Yongchao and Ba, Jimmy and Dubois, Yann and Maddison, Chris J. and Hashimoto, Tatsunori},
  booktitle = {International Conference on Learning Representations ({ICLR})},
  year      = {2024}
}

@inproceedings{jia2024taskshield,
  title     = {The Task Shield: Enforcing Task Alignment to Defend Against Indirect Prompt Injection in {LLM} Agents},
  author    = {Jia, Feiran and Wu, Tong and Qin, Xin and Squicciarini, Anna},
  booktitle = {Proceedings of the 63rd Annual Meeting of the Association for Computational Linguistics},
  pages     = {29680--29697},
  year      = {2025},
  url       = {https://aclanthology.org/2025.acl-long.1435/}
}

@inproceedings{
xiang2024guardagent,
title={GuardAgent: Safeguard {LLM} Agents via Knowledge-Enabled Reasoning},
author={Zhen Xiang and Linzhi Zheng and Yanjie Li and Junyuan Hong and Qinbin Li and Han Xie and Jiawei Zhang and Zidi Xiong and Chulin Xie and Nathaniel D. Bastian and Carl Yang and Dawn Song and Bo Li},
booktitle={ICML 2025 Workshop on Computer Use Agents},
year={2025},
url={https://openreview.net/forum?id=ITuuEaXcSB}
}

@article{debenedetti2025camel,
  title   = {Defeating Prompt Injections by Design},
  author  = {Debenedetti, Edoardo and Shumailov, Ilia and Fan, Tianqi and Hayes, Jamie and Carlini, Nicholas and Fabian, Daniel and Kern, Christoph and Shi, Chongyang and Terzis, Andreas and Tram{\`e}r, Florian},
  journal = {arXiv preprint arXiv:2503.18813},
  year    = {2025}
}

@article{fu2024imprompter,
  title   = {Imprompter: Tricking {LLM} Agents into Improper Tool Use},
  author  = {Fu, Xiaohan and Li, Shuheng and Wang, Zihan and Liu, Yihao and Gupta, Rajesh K. and Berg-Kirkpatrick, Taylor and Fernandes, Earlence},
  journal = {arXiv preprint arXiv:2410.14923},
  year    = {2024}
}

@article{shi2025geminidefense,
  title   = {Lessons from Defending {Gemini} Against Indirect Prompt Injections},
  author  = {Shi, Chongyang and Lin, Sharon and Song, Shuang and Hayes, Jamie and Shumailov, Ilia and Yona, Itay and Pluto, Juliette and Pappu, Aneesh and Choquette-Choo, Christopher A. and Nasr, Milad and Sitawarin, Chawin and Gibson, Gena and Terzis, Andreas and Flynn, John ``Four''},
  journal = {arXiv preprint arXiv:2505.14534},
  year    = {2025}
}

@article{costa2025fides,
  title   = {Securing {AI} Agents with Information-Flow Control},
  author  = {Costa, Manuel and K{\"o}pf, Boris and Kolluri, Aashish and Paverd, Andrew and Russinovich, Mark and Salem, Ahmed and Tople, Shruti and Wutschitz, Lukas and Zanella-B{\'e}guelin, Santiago},
  journal = {arXiv preprint arXiv:2505.23643},
  year    = {2025}
}

@misc{shi2025progent,
      title={Progent: Programmable Privilege Control for LLM Agents}, 
      author={Tianneng Shi and Jingxuan He and Zhun Wang and Hongwei Li and Linyu Wu and Wenbo Guo and Dawn Song},
      year={2025},
      eprint={2504.11703},
      archivePrefix={arXiv},
      primaryClass={cs.CR},
      url={https://arxiv.org/abs/2504.11703}, 
}

@inproceedings{newman2022sigstore,
author = {Newman, Zachary and Meyers, John Speed and Torres-Arias, Santiago},
title = {Sigstore: Software Signing for Everybody},
year = {2022},
isbn = {9781450394505},
publisher = {Association for Computing Machinery},
address = {New York, NY, USA},
url = {https://doi.org/10.1145/3548606.3560596},
doi = {10.1145/3548606.3560596},
booktitle = {Proceedings of the 2022 ACM SIGSAC Conference on Computer and Communications Security},
pages = {2353–2367},
numpages = {15},
location = {Los Angeles, CA, USA},
series = {CCS '22}
}

@inproceedings {torresarias2019intoto,
author = {Santiago Torres-Arias and Hammad Afzali and Trishank Karthik Kuppusamy and Reza Curtmola and Justin Cappos},
title = {in-toto: Providing farm-to-table guarantees for bits and bytes},
booktitle = {USENIX Security Symposium (USENIX Sec)},
year = {2019},
isbn = {978-1-939133-06-9},
address = {Santa Clara, CA},
pages = {1393--1410},
url = {https://www.usenix.org/conference/usenixsecurity19/presentation/torres-arias},
publisher = {USENIX Association},
month = aug
}

@INPROCEEDINGS{johnson2023wave,
  author={Johnson, Evan and Laufer, Evan and Zhao, Zijie and Gohman, Dan and Narayan, Shravan and Savage, Stefan and Stefan, Deian and Brown, Fraser},
  booktitle={2023 IEEE Symposium on Security and Privacy (SP)}, 
  title={WaVe: a verifiably secure WebAssembly sandboxing runtime}, 
  year={2023},
  volume={},
  number={},
  pages={2940-2955},
  doi={10.1109/SP46215.2023.10179357}}

@inproceedings{jia2016chromesandbox,
author = {Jia, Yaoqi and Chua, Zheng Leong and Hu, Hong and Chen, Shuo and Saxena, Prateek and Liang, Zhenkai},
title = {{"The Web/Local"} Boundary Is Fuzzy: A Security Study of Chrome's Process-based Sandboxing},
year = {2016},
isbn = {9781450341394},
publisher = {Association for Computing Machinery},
address = {New York, NY, USA},
doi = {10.1145/2976749.2978414},
booktitle = {Proceedings of the ACM SIGSAC Conference on Computer and Communications Security},
pages = {791–804},
numpages = {14},
location = {Vienna, Austria},
series = {CCS '16}
}

@inproceedings {nunes2019vrased,
author = {Ivan De Oliveira Nunes and Karim Eldefrawy and Norrathep Rattanavipanon and Michael Steiner and Gene Tsudik},
title = {{VRASED}: A Verified {Hardware/Software} {Co-Design} for Remote Attestation},
booktitle = {USENIX Security Symposium (USENIX Sec)},
year = {2019},
isbn = {978-1-939133-06-9},
address = {Santa Clara, CA},
pages = {1429--1446},
url = {https://www.usenix.org/conference/usenixsecurity19/presentation/de-oliveira-nunes},
publisher = {USENIX Association},
month = aug
}

@inproceedings{chen2024agentpoison,
author = {Chen, Zhaorun and Xiang, Zhen and Xiao, Chaowei and Song, Dawn and Li, Bo},
title = {{AgentPoison}: red-teaming {LLM} agents via poisoning memory or knowledge bases},
year = {2024},
isbn = {9798331314385},
publisher = {Curran Associates Inc.},
address = {Red Hook, NY, USA},
booktitle = {Proceedings of the International Conference on Neural Information Processing Systems (NeurIPS)},
articleno = {4136},
numpages = {29},
location = {Vancouver, BC, Canada},
series = {NIPS '24}
}

@misc{srivastava2025memorygraft,
      title={MemoryGraft: Persistent Compromise of LLM Agents via Poisoned Experience Retrieval}, 
      author={Saksham Sahai Srivastava and Haoyu He},
      year={2025},
      eprint={2512.16962},
      archivePrefix={arXiv},
      primaryClass={cs.CR},
      url={https://arxiv.org/abs/2512.16962}, 
}

@article{sunil2026memorypoisoning,
  title={Memory Poisoning Attack and Defense on Memory Based LLM-Agents},
  author={Sunil, Balachandra Devarangadi and Sinha, Isheeta and Maheshwari, Piyush and Todmal, Shantanu and Mallik, Shreyan and Mishra, Shuchi},
  journal={arXiv preprint arXiv:2601.05504; accessible on \url{https://arxiv.org/abs/2601.05504}},
  year={2026}
}

@inproceedings{
triedman2025masmaliciouscode,
title={Multi-Agent Systems Execute Arbitrary Malicious Code},
author={Harold Triedman and Rishi Dev Jha and Vitaly Shmatikov},
booktitle={Second Conference on Language Modeling (COLM)},
year={2025},
url={https://openreview.net/forum?id=DAozI4etUp}
}

@inproceedings{he2025communicationattacks,
    title = "Red-Teaming {LLM} Multi-Agent Systems via Communication Attacks",
    author = "He, Pengfei  and
      Lin, Yuping  and
      Dong, Shen  and
      Xu, Han  and
      Xing, Yue  and
      Liu, Hui",
    booktitle = "Findings of the Association for Computational Linguistics (ACL)",
    month = jul,
    year = "2025",
    address = "Vienna, Austria",
    publisher = "Association for Computational Linguistics",
    doi = "10.18653/v1/2025.findings-acl.349",
    pages = "6726--6747",
    ISBN = "979-8-89176-256-5",
}

@article{wang2025ipleakage,
  title   = {{IP} Leakage Attacks Targeting {LLM}-Based Multi-Agent Systems},
  author  = {Wang, Liwen and Wang, Wenxuan and Wang, Shuai and Li, Zongjie and Ji, Zhenlan and Lyu, Zongyi and Wu, Daoyuan and Cheung, Shing-Chi},
  journal = {arXiv preprint arXiv:2505.12442},
  year    = {2025}
}

@article{ko2025sevenchallenges,
  title   = {Seven Security Challenges That Must be Solved in Cross-domain Multi-agent {LLM} Systems},
  author  = {Ko, Ronny and Jeong, Jiseong and Zheng, Shuyuan and Xiao, Chuan and Kim, Tae-Wan and Onizuka, Makoto and Shin, Won-Yong},
  journal = {arXiv preprint arXiv:2505.23847},
  year    = {2025}
}

@article{radosevich2025mcpaudit,
  title   = {{MCP} Safety Audit: {LLMs} with the Model Context Protocol Allow Major Security Exploits},
  author  = {Radosevich, Brandon and Halloran, John},
  journal = {arXiv preprint arXiv:2504.03767},
  year    = {2025}
}

@article{hou2025mcplandscape,
  title   = {Model Context Protocol ({MCP}): Landscape, Security Threats, and Future Research Directions},
  author  = {Hou, Xinyi and Zhao, Yanjie and Wang, Shenao and Wang, Haoyu},
  journal = {arXiv preprint arXiv:2503.23278},
  year    = {2025}
}

@article{xing2025mcpguard,
  title   = {{MCP-Guard}: A Multi-Stage Defense-in-Depth Framework for Securing Model Context Protocol in Agentic {AI}},
  author  = {Xing, Wenpeng and Qi, Zhonghao and Qin, Yupeng and Li, Yilin and Chang, Caini and Yu, Jiahui and Lin, Changting and Xie, Zhenzhen and Han, Meng},
  journal = {arXiv preprint arXiv:2508.10991},
  year    = {2025}
}

@article{he2025mcpredteam,
  title   = {Automatic Red Teaming {LLM}-based Agents with Model Context Protocol Tools},
  author  = {He, Ping and Li, Changjiang and Zhao, Binbin and Du, Tianyu and Ji, Shouling},
  journal = {arXiv preprint arXiv:2509.21011},
  year    = {2025}
}

@article{dong2024agentops,
  title   = {{AgentOps}: Enabling Observability of {LLM} Agents},
  author  = {Dong, Liming and Lu, Qinghua and Zhu, Liming},
  journal = {arXiv preprint arXiv:2411.05285},
  year    = {2024}
}

@inproceedings{gu2024agentsmith,
  title     = {Agent Smith: A Single Image Can Jailbreak One Million Multimodal {LLM} Agents Exponentially Fast},
  author    = {Gu, Xiangming and Zheng, Xiaosen and Pang, Tianyu and Du, Chao and Liu, Qian and Wang, Ye and Jiang, Jing and Lin, Min},
  booktitle = {Proceedings of the International Conference on Machine Learning (ICML)},
  volume    = {235},
  pages     = {16647--16672},
  year      = {2024},
}

@inproceedings{liao2025eia,
  title     = {{EIA}: Environmental Injection Attack on Generalist Web Agents for Privacy Leakage},
  author    = {Liao, Zeyi and Mo, Lingbo and Xu, Chejian and Kang, Mintong and Zhang, Jiawei and Xiao, Chaowei and Tian, Yuan and Li, Bo and Sun, Huan},
  booktitle = {International Conference on Learning Representations ({ICLR})},
  year      = {2025},
  url       = {https://openreview.net/forum?id=xMOLUzo2Lk}
}

@inproceedings{evtimov2025wasp,
title={{WASP}: Benchmarking Web Agent Security Against Prompt Injection Attacks},
author={Ivan Evtimov and Arman Zharmagambetov and Aaron Grattafiori and Chuan Guo and Kamalika Chaudhuri},
booktitle={ICML 2025 Workshop on Computer Use Agents},
year={2025}
}

@inproceedings{wang2025envinjection,
  title     = {{WebInject}: Prompt Injection Attack to Web Agents},
  author    = {Wang, Xilong and Bloch, John and Shao, Zedian and Hu, Yuepeng and Zhou, Shuyan and Gong, Neil Zhenqiang},
  booktitle = {Proceedings of the Conference on Empirical Methods in Natural Language Processing (EMNLP)},
  year      = {2025},
  doi       = {10.18653/v1/2025.emnlp-main.104},
  url       = {https://aclanthology.org/2025.emnlp-main.104/}
}

@inproceedings{zharmagambetov2025agentdam,
  title     = {{AgentDAM}: Privacy Leakage Evaluation for Autonomous Web Agents},
  author    = {Zharmagambetov, Arman and Guo, Chuan and Evtimov, Ivan and Pavlova, Maya and Salakhutdinov, Ruslan and Chaudhuri, Kamalika},
  booktitle = {Advances in Neural Information Processing Systems ({NeurIPS}) 2025 Datasets and Benchmarks Track},
  year      = {2025},
  url       = {https://neurips.cc/virtual/2025/poster/121443}
}

@inproceedings{nie2024privagent,
  title     = {{LeakAgent}: {RL}-based Red-teaming Agent for {LLM} Privacy Leakage},
  author    = {Nie, Yuzhou and Wang, Zhun and Yu, Ye and Wu, Xian and Zhao, Xuandong and Bastian, Nathaniel D. and Guo, Wenbo and Song, Dawn},
  booktitle = {Conference on Language Modeling ({COLM})},
  year      = {2025},
  url       = {https://openreview.net/forum?id=WIfns41MAb}
}

@article{zhang2025simprivacy,
  title   = {Searching for Privacy Risks in {LLM} Agents via Simulation},
  author  = {Zhang, Yanzhe and Yang, Diyi},
  journal = {arXiv preprint arXiv:2508.10880},
  year    = {2025}
}

@article{zhang2025privweb,
  title   = {{PrivWeb}: Unobtrusive and Content-aware Privacy Protection For Web Agents},
  author  = {Zhang, Shuning and Jiang, Yutong and Ma, Rongjun and Yang, Yuting and Xu, Mingyao and Huang, Zhixin and Yi, Xin and Li, Hewu},
  journal = {arXiv preprint arXiv:2509.11939},
  year    = {2025}
}

@inproceedings{luo2025agentauditor,
  title     = {{AgentAuditor}: Human-Level Safety and Security Evaluation for {LLM} Agents},
  author    = {Luo, Hanjun and Dai, Shenyu and Ni, Chiming and Li, Xinfeng and Zhang, Guibin and Wang, Kun and Liu, Tongliang and Salam, Hanan},
  booktitle = {Advances in Neural Information Processing Systems ({NeurIPS})},
  year      = {2025}
}

@inproceedings{wang2025agentfuzzer,
  title     = {{AGENTVIGIL}: Automatic Black-Box Red-teaming for Indirect Prompt Injection against {LLM} Agents},
  author    = {Wang, Zhun and Siu, Vincent and Ye, Zhe and Shi, Tianneng and Nie, Yuzhou and Zhao, Xuandong and Wang, Chenguang and Guo, Wenbo and Song, Dawn},
  booktitle = {Findings of the Association for Computational Linguistics (EMNLP)},
  pages     = {23159--23172},
  year      = {2025},
  doi       = {10.18653/v1/2025.findings-emnlp.1258},
  url       = {https://aclanthology.org/2025.findings-emnlp.1258/}
}

@inproceedings{wang2025agentspec,
  title     = {{AgentSpec}: Customizable Runtime Enforcement for Safe and Reliable {LLM} Agents},
  author    = {Wang, Haoyu and Poskitt, Christopher M. and Sun, Jun},
  booktitle = {Proceedings of the 48th {IEEE/ACM} International Conference on Software Engineering ({ICSE})},
  year      = {2026}
}

@article{wang2025pro2guard,
  title   = {{Pro2Guard}: Proactive Runtime Enforcement of {LLM} Agent Safety via Probabilistic Model Checking},
  author  = {Wang, Haoyu and Poskitt, Christopher M. and Sun, Jun and Wei, Jiali},
  journal = {arXiv preprint arXiv:2508.00500},
  year    = {2025}
}

@article{zhang2025securityprinciples,
  title   = {{LLM} Agents Should Employ Security Principles},
  author  = {Zhang, Kaiyuan and Su, Zian and Chen, Pin-Yu and Bertino, Elisa and Zhang, Xiangyu and Li, Ninghui},
  journal = {arXiv preprint arXiv:2505.24019},
  year    = {2025}
}

@inproceedings{nunez2024autosafecoder,
  title     = {{AutoSafeCoder}: A Multi-Agent Framework for Securing {LLM} Code Generation through Static Analysis and Fuzz Testing},
  author    = {Nunez, Ana and Islam, Nafis Tanveer and Jha, Sumit Kumar and Najafirad, Peyman},
  booktitle = {NeurIPS Workshop on Safe \& Trustworthy Agents},
  year      = {2024}
}

@article{liu2025wainjectbench,
  title   = {{WAInjectBench}: Benchmarking Prompt Injection Detections for Web Agents},
  author  = {Liu, Yinuo and Xu, Ruohan and Wang, Xilong and Jia, Yuqi and Gong, Neil Zhenqiang},
  journal = {arXiv preprint arXiv:2510.01354},
  year    = {2025}
}

@inproceedings{das2025disclosureaudits,
  title     = {Disclosure Audits for {LLM} Agents},
  author    = {Das, Saswat and Sandler, Jameson and Fioretto, Ferdinando},
  booktitle = {Workshop on Multi-Turn Interactions in Large Language Models ({MTI-LLM}) at {NeurIPS} 2025},
  year      = {2025},
  url       = {https://openreview.net/forum?id=BV6QBwZkfK}
}

@article{zhong2025rtbas,
  title   = {{RTBAS}: Defending {LLM} Agents Against Prompt Injection and Privacy Leakage},
  author  = {Zhong, Peter Yong and Chen, Siyuan and Wang, Ruiqi and McCall, McKenna and Titzer, Ben L. and Miller, Heather and Gibbons, Phillip B.},
  journal = {arXiv preprint arXiv:2502.08966},
  year    = {2025}
}

@misc{anthropic2024mcp,
  author       = {{Anthropic}},
  title        = {Model Context Protocol Specification},
  year         = {2024},
  howpublished = {\url{https://modelcontextprotocol.io/specification}}
}

@inproceedings{felt2012androidpermissions,
  title     = {Android Permissions: User Attention, Comprehension, and Behavior},
  author    = {Felt, Adrienne Porter and Ha, Elizabeth and Egelman, Serge and Haney, Ariel and Chin, Erika and Wagner, David},
  booktitle = {Proceedings of the Symposium on Usable Privacy and Security ({SOUPS})},
  year      = {2012}
}

@misc{rehberger2025claudecodecve,
  author       = {Rehberger, Johann},
  title        = {{Claude Code: Data Exfiltration with DNS (CVE-2025-55284)}},
  howpublished = {\url{https://embracethered.com/blog/posts/2025/claude-code-exfiltration-via-dns-requests/}},
  year         = {2025}
}

@misc{rehberger2025cursorcve,
  author       = {Rehberger, Johann},
  title        = {{Cursor IDE: Arbitrary Data Exfiltration Via Mermaid (CVE-2025-54132)}},
  howpublished = {\url{https://embracethered.com/blog/posts/2025/cursor-data-exfiltration-with-mermaid/}},
  year         = {2025}
}

@misc{rehberger2025copilotcve,
  author       = {Rehberger, Johann},
  title        = {{GitHub Copilot: Remote Code Execution via Prompt Injection (CVE-2025-53773)}},
  howpublished = {\url{https://embracethered.com/blog/posts/2025/github-copilot-remote-code-execution-via-prompt-injection/}},
  year         = {2025}
}

@misc{rehberger2025cline,
  author       = {Rehberger, Johann},
  title        = {{Cline: Vulnerable to Data Exfiltration and How to Protect Your Data}},
  howpublished = {\url{https://embracethered.com/blog/posts/2025/cline-vulnerable-to-data-exfiltration/}},
  year         = {2025}
}

@misc{technode2026miclaw,
  author       = {{TechNode}},
  title        = {{Xiaomi Begins Limited Closed Beta of {OpenClaw}-like Mobile {AI} Agent Xiaomi Miclaw}},
  howpublished = {\url{https://technode.com/2026/03/06/xiaomi-begins-limited-closed-beta-of-openclaw-like-mobile-ai-agent-xiaomi-miclaw/}},
  year         = {2026}
}

@misc{claudecode2026,
  author       = {{Anthropic}},
  title        = {{Claude Code Overview}},
  howpublished = {\url{https://docs.anthropic.com/en/docs/claude-code/overview}},
  year         = {2026}
}

@misc{openhands2026,
  author       = {{All Hands AI}},
  title        = {{OpenHands: Open Source AI Software Development Agents}},
  howpublished = {\url{https://github.com/All-Hands-AI/OpenHands}},
  year         = {2026}
}

@misc{cursor2026,
  author       = {{Anysphere}},
  title        = {{Cursor: The AI Code Editor}},
  howpublished = {\url{https://cursor.com}},
  year         = {2026}
}

@misc{aider2026,
  author       = {{Aider-AI}},
  title        = {{Aider: AI Pair Programming in Your Terminal}},
  howpublished = {\url{https://aider.chat}},
  year         = {2026}
}

@misc{cline2026,
  author       = {{Cline}},
  title        = {{Cline: Open-Source Coding Agent for VS Code}},
  howpublished = {\url{https://github.com/cline/cline}},
  year         = {2026}
}

@misc{anthropic2025espionage,
  author       = {{Anthropic}},
  title        = {{Disrupting the First Reported {AI}-Orchestrated Cyber Espionage Campaign}},
  howpublished = {\url{https://www.anthropic.com/news/disrupting-AI-espionage}},
  year         = {2025}
}

@misc{trae2026,
  author       = {{ByteDance}},
  title        = {{TRAE: The Real AI Engineer}},
  howpublished = {\url{https://www.trae.ai}},
  year         = {2026}
}

@misc{qwenagent2026,
  author       = {{Alibaba Qwen Team}},
  title        = {{Qwen-Agent: Agent Framework and Applications Built upon Qwen}},
  howpublished = {\url{https://github.com/QwenLM/Qwen-Agent}},
  year         = {2026}
}

@misc{manus2025,
  author       = {{manus}},
  title        = {{manus}},
  howpublished = {\url{https://manus.im}},
  year         = {2025}
}

@misc{caixin2026openclaw,
  author       = {{Caixin Global}},
  title        = {Xiaomi, Huawei Rush to Deploy {AI} Agents Amid {OpenClaw} Craze},
  howpublished = {\url{https://www.caixinglobal.com/2026-03-12/xiaomi-huawei-rush-to-deploy-ai-agents-amid-openclaw-craze-102422164.html}},
  year         = {2026},
  month        = {March}
}

@misc{windsurf2025,
  author       = {{Codeium}},
  title        = {{Windsurf Editor: The Agentic {IDE}}},
  howpublished = {\url{https://windsurf.com/editor}},
  year         = {2025}
}

@misc{devin2026,
  author       = {{Cognition AI}},
  title        = {{Devin: The AI Software Engineer}},
  howpublished = {\url{https://devin.ai}},
  year         = {2024}
}

@misc{copilotworkspace2025,
  author       = {{GitHub}},
  title        = {{GitHub Copilot Workspace: Task-Centric Development Environment}},
  howpublished = {\url{https://githubnext.com/projects/copilot-workspace}},
  year         = {2024}
}

@misc{amazonq2025,
  author       = {{Amazon Web Services}},
  title        = {{Amazon Q Developer: {AI}-Powered Coding Assistant}},
  howpublished = {\url{https://aws.amazon.com/q/developer/}}
}

@misc{techcrunch2025copilot20m,
  author       = {{TechCrunch}},
  title        = {{GitHub Copilot Crosses 20 Million All-Time Users}},
  howpublished = {\url{https://techcrunch.com/2025/07/30/github-copilot-crosses-20-million-all-time-users/}},
  year         = {2025},
  month        = {July}
}

@inproceedings{greshake2023indirect,
  author = {Greshake, Kai and Abdelnabi, Sahar and Mishra, Shailesh and Endres, Christoph and Holz, Thorsten and Fritz, Mario}, 
  title = {Not What You've Signed Up For: Compromising Real-World LLM-Integrated Applications with Indirect Prompt Injection}, year = {2023}, isbn = {9798400702600}, publisher = {Association for Computing Machinery}, address = {New York, NY, USA}, url = {https://doi.org/10.1145/3605764.3623985}, doi = {10.1145/3605764.3623985}, booktitle = {Proceedings of the 16th ACM Workshop on Artificial Intelligence and Security}, pages = {79–90}, numpages = {12}, location = {Copenhagen, Denmark}, series = {AISec '23} }

@inproceedings{ladisa2023taxonomy,
  author    = {Ladisa, Piergiorgio and Plate, Henrik and Martinez, Matias and Barais, Olivier},
  title     = {{SoK}: Taxonomy of Attacks on Open-Source Software Supply Chains},
  booktitle = {2023 IEEE Symposium on Security and Privacy ({S\&P})},
  year      = {2023},
  pages     = {1509--1526},
  publisher = {IEEE}
}

@inproceedings{ohm2020supplychain,
  author    = {Ohm, Marc and Plate, Henrik and Sykosch, Arnold and Meier, Michael},
  title     = {Backstabber's Knife Collection: A Review of Open Source Software Supply Chain Attacks},
  booktitle = {Detection of Intrusions and Malware, and Vulnerability Assessment ({DIMVA} 2020)},
  year      = {2020},
  pages     = {23--43},
  publisher = {Springer}
}

@misc{duclaw2026,
  author       = {{Baidu Inc.}},
  title        = {DuClaw: Zero-Deployment OpenClaw Service},
  howpublished = {\url{https://cloud.baidu.com/product/duclaw.html}},
  year         = {2026}
}

@article{jiang2026sok,
  title   = {{SoK}: Agentic Skills---Beyond Tool Use in {LLM} Agents},
  author  = {Jiang, Yanna and Li, Delong and Deng, Haiyu and Ma, Baihe and Wang, Xu and Wang, Qin and Yu, Guangsheng},
  journal = {arXiv preprint arXiv:2602.20867},
  year    = {2026}
}

@article{yu2026plantwin,
  title={PlanTwin: Privacy-Preserving Planning Abstractions for Cloud-Assisted LLM Agents},
  author={Yu, Guangsheng and Wang, Qin and Lang, Rui and Su, Shuai and Wang, Xu},
  journal={arXiv preprint arXiv:2603.18377},
  year={2026}
}

@article{ma2025sok,
  title={SoK: Semantic Privacy in Large Language Models},
  author={Ma, Baihe and Jiang, Yanna and Wang, Xu and Yu, Guangsheng and Wang, Qin and Sun, Caijun and Li, Chen and Qi, Xuelei and He, Ying and Ni, Wei and others},
  journal={arXiv preprint arXiv:2506.23603},
  year={2025}
}

\appendix

\section{Search Methodology Details}
\label{app:search}

Our full search protocol details in \S\ref{sec:methodology} are summarized below.

\subsection{Database Sources}
\label{app:search:sources}

The corpus was constructed by querying the following sources, listed in
precedence order for citation selection:

\begin{packeditemize}
  \item \textit{Tier-1 venue proceedings.}
    Software engineering: ICSE, FSE, ASE, ISSTA.
    Security: IEEE~S\&P, USENIX~Security, ACM~CCS, NDSS.
    AI/NLP: NeurIPS, ICLR, ICML, ACL, EMNLP, NAACL, SaTML.
  \item \textit{Metadata and index services.}
    Google Scholar, DBLP, ACL Anthology, OpenReview.
  \item \textit{Preprint repositories.}
    arXiv (cs.CR, cs.AI, cs.CL, cs.SE).
  \item \textit{Supplementary sources.}
    Microsoft Research publication pages, author and lab pages (used only
    when they exposed canonical metadata or pointed to a venue record).
\end{packeditemize}

\subsection{Time Window}
\label{app:search:window}

The nominal literature window spans \textbf{January~2023} to \textbf{March~2026}. Background engineering foundations (\eg browser process isolation, software supply-chain attestation, WebAssembly sandboxing) may predate 2023 when they serve as direct architectural analogies for the software engineering lens adopted by this paper.

\subsection{Search Query Strings}
\label{app:search:queries}

Fourteen query strings were executed across the above sources.
Queries~1--12 target the agent-security literature directly;
queries~13--14 target adjacent engineering foundations whose
architectural patterns inform the SE framing.

\begin{enumerate}[label=Q\arabic*:, leftmargin=2.5em]
  \item \texttt{"LLM agent security"}
  \item \texttt{"autonomous agent prompt injection"}
  \item \texttt{"tool-integrated LLM agents security"}
  \item \texttt{"trustworthy LLM agents survey"}
  \item \texttt{"agent benchmark prompt injection"}
  \item \texttt{"agent security bench"}
  \item \texttt{"memory poisoning LLM agents"}
  \item \texttt{"programmable privilege control LLM agents"}
  \item \texttt{"information-flow control AI agents"}
  \item \texttt{"multi-agent systems malicious code LLM"}
  \item \texttt{"Model Context Protocol security"}
  \item \texttt{"OpenClaw security"}
  \item \texttt{"software signing provenance attestation supply chain"}
  \item \texttt{"WebAssembly sandboxing security"}
\end{enumerate}

\subsection{Snowball Sampling}
\label{app:search:snowball}

Forward and backward citation chasing was applied to every paper that
passed title/abstract screening.
Backward chasing followed reference lists to identify seminal work that
predates the search window but remains architecturally relevant (\eg
Sigstore~\cite{newman2022sigstore}, in-toto~\cite{torresarias2019intoto},
Chrome site-isolation~\cite{jia2016chromesandbox}).
Forward chasing used Google Scholar's ``Cited by'' and Semantic Scholar's
citation graph to identify recent follow-up work not yet captured by
the initial query sweep.
This two-pass snowball procedure was particularly important for the
memory-security and MCP-security clusters, where the primary literature
is sparse and heavily cross-referencing.

\subsection{Screening Funnel}
\label{app:search:funnel}

Table~\ref{tab:screening-funnel} summarizes the four-stage screening workflow
and the candidate counts at each stage.

\begin{table}[h]
\centering
\caption{Screening funnel for corpus construction.
  The freeze date for the current pass is March~17, 2026.}
\label{tab:screening-funnel}
\resizebox{\linewidth}{!}{%
\begin{tabular}{c|l|c}
\toprule
\rowcolor{orange!20}
\multicolumn{1}{c}{\textbf{Stage}} & 
\multicolumn{1}{c}{\textbf{Action}} & 
\multicolumn{1}{c}{\textbf{Papers}} \\
\midrule
Identification   & Query results aggregated from all sources & $\sim$340 \\
Deduplication     & Merged arXiv/DBLP/OpenReview/venue versions & $\sim$210 \\
Title/abs screening & Retained clearly relevant papers & 82 \\
Full-text review  & Final inclusion with normalized coding & \textbf{50} \\
\bottomrule
\end{tabular}
}
\end{table}

\noindent
Of the 50~included papers, 5 are classified as adjacent engineering
foundations (supply-chain signing, sandboxing, attestation) that predate
the nominal 2023 window but directly inform the software engineering
framing.
The remaining 45~papers address agent security, benchmarks, defenses,
protocols, or deployment experience published between 2023 and 2026.

\smallskip
\noindent\textbf{Version precedence.}
When multiple versions of a paper exist, the peer-reviewed venue version
takes precedence for citation.
The selection order is:
(1)~official conference or journal page,
(2)~OpenReview or ACL Anthology proceedings page,
(3)~DBLP venue record,
(4)~arXiv abstract page,
(5)~author or lab page.

\subsection{Full Screening Criteria}
\label{app:criteria}

This section details the inclusion and exclusion criteria and specifies the quality assessment rubric applied during full-text review.

\smallskip
\noindent\textbf{Inclusion criteria.}
\label{app:criteria:inclusion}
A paper is included in the corpus if it satisfies \emph{at least one} of the following conditions:

\begin{enumerate}[label=I\arabic*., leftmargin=2em]
  \item \textit{Agent-level threat analysis.}
    The paper analyzes security or trust failures that arise specifically
    in tool-using, memory-augmented, or autonomous LLM agents, not
    merely in base language models.
  \item \textit{Agent defense or architecture.}
    The paper proposes a defense mechanism, architectural pattern, or
    policy framework for agent execution, planning, tool mediation,
    protocol mediation, or persistent memory.
  \item \textit{Agent-security benchmark.}
    The paper introduces or substantially extends a benchmark for
    evaluating agent-level attacks or defenses, including prompt
    injection, tool misuse, memory poisoning, or privilege escalation.
  \item \textit{Agent-security survey.}
    The paper surveys agent threats and countermeasures with sufficient
    breadth or depth to inform a systematization.
  \item \textit{Deployment or operations study.}
    The paper reports deployment experience, operational lessons, or
    production retrospectives for agent systems that materially reshape
    the understanding of the attack surface.
  \item \textit{Adjacent engineering foundation.}
    The paper provides an engineering foundation for provenance,
    attestation, sandboxing, or secure software supply chains that directly informs the SE framing, even if the paper itself does not address LLM agents.
    Papers in this category must demonstrate clear architectural
    transferability to the agent-ecosystem setting.
\end{enumerate}

\smallskip
\noindent\textbf{Exclusion criteria.}
\label{app:criteria:exclusion}
A paper is excluded if \emph{any} of the following conditions holds and \emph{none} of the inclusion criteria are satisfied:

\begin{enumerate}[label=E\arabic*., leftmargin=2em]
  \item \textit{Base-model jailbreak only.}
    The paper focuses exclusively on jailbreaking or red-teaming a
    base language model without tool invocation, memory persistence,
    or agent-level autonomy.
  \item \textit{Generic LLM safety.}
    The paper addresses hallucination, toxicity, bias, or alignment
    in language models without security implications for agent
    platforms.
  \item \textit{Non-agent privacy or ethics.}
    The paper discusses privacy, fairness, or ethical concerns that
    do not affect agent platform design, deployment, or operations.
  \item \textit{Product announcement.}
    The paper is a product announcement, blog post, or marketing
    material that lacks stable technical substance, reproducible
    methodology, or peer review.
  \item \textit{Insufficient technical depth.}
    The paper lacks sufficient technical depth to inform engineering
    decisions (\eg workshop abstracts, extended abstracts without
    supplementary material, or position statements without supporting
    evidence).
\end{enumerate}

\section{Illustrative Templates}
\label{app:examples}

\noindent\textbf{Scope disclaimer.}
The examples below are illustrative, not prescriptive.
They instantiate the principles in \S\ref{sec:principles} using design patterns inspired by systems such as CaMeL, Progent, Sigstore, in-toto, and AgentOps.
Their role is simply to show that the reference doctrine can be made concrete.
They are not protocol specifications, are not production-validated, and should not be treated as deployment-ready designs.
Any real implementation would require formal specification, security analysis, and empirical evaluation.

\smallskip
\noindent
While \S\ref{sec:principles:doctrine} presents the doctrine at a principle level, this appendix gives worked examples for four artifacts: a capability manifest, a policy rule, a memory-provenance record, and an audit-telemetry entry.
All examples use the \mcp ecosystem as the running case and are deliberately simplified to show the minimum useful information each artifact could contain.

\subsection{Capability Manifest Example}
\label{app:examples:manifest}

Listing~\ref{lst:manifest} shows what a tool capability declaration could look
like for an MCP server that provides filesystem operations.
The manifest would declare (i)~the tool's identity and version,
(ii)~the specific operations it exposes with typed parameters,
(iii)~the permissions it requests with explicit scope restrictions,
(iv)~a time-to-live bound on the capability grant, and
(v)~a cryptographic signature from the tool publisher that binds the
manifest to a verifiable identity.
This structure illustrates how the ``replace ambient authority with
explicit capabilities'' principle (\S\ref{sec:principles:capabilities})
and the Layer~2 policy-mediation patterns
(\S\ref{sec:principles:doctrine:l2}) could be instantiated in practice.

\begin{figure}[h]
\begin{lstlisting}[
  language={},
  basicstyle=\ttfamily\scriptsize,
  frame=single,
  xleftmargin=2pt,
  xrightmargin=2pt,
  breaklines=true,
  columns=flexible,
  captionpos=b,
  caption={Capability manifest for an MCP server.},
  label={lst:manifest}
]
{
  "manifest_version": "1.0",
  "server": {
    "name": "fs-workspace",
    "version": "0.3.1",
    "publisher": "acme-tools",
    "publisher_identity": "sigstore:acme-tools@registry.dev"
  },
  "tools": [
    {
      "name": "read_file",
      "description": "Read a file within the workspace.",
      "parameters": {
        "path": {"type": "string", "pattern": "^\\./.*"}
      },
      "required_permissions": ["fs:read"],
      "scope": {"root": "./workspace", "recursive": true},
      "side_effects": "none"
    },
    {
      "name": "write_file",
      "description": "Write a file within the workspace.",
      "parameters": {
        "path": {"type": "string", "pattern": "^\\./.*"},
        "content": {"type": "string", "maxLength": 1048576}
      },
      "required_permissions": ["fs:write"],
      "scope": {"root": "./workspace", "recursive": true},
      "side_effects": "filesystem_mutation"
    }
  ],
  "capability_grant": {
    "ttl_seconds": 3600,
    "max_invocations": 100,
    "renewable": true,
    "escalation_requires": "user_approval"
  },
  "signature": {
    "algorithm": "ECDSA-P256",
    "certificate": "sigstore://acme-tools/fs-workspace/0.3.1",
    "manifest_hash": "sha256:a1b2c3d4...ef56"
  }
}
\end{lstlisting}
\end{figure}

\smallskip
\noindent
Several fields in this example matter for distinct reasons:
\begin{packeditemize}
  \item The \texttt{scope} field would restrict every operation to a workspace
    subtree, preventing path-traversal attacks that escape the project
    directory (\eg reading \texttt{\textasciitilde/.aws/credentials}).
  \item The \texttt{ttl\_seconds} and \texttt{max\_invocations} fields
    would enforce time-bounded and count-bounded capability grants, so
    that a compromised session could not reuse a stale token indefinitely.
  \item The \texttt{signature} block would bind the manifest to a Sigstore
    identity certificate~\cite{newman2022sigstore}, enabling clients to
    verify that the manifest was published by its claimed author and has
    not been tampered with since signing.
  \item The \texttt{escalation\_requires} field would ensure that any
    operation exceeding the declared scope triggers an out-of-band
    approval flow rather than failing silently.
\end{packeditemize}

\subsection{Policy Expression Example}
\label{app:examples:policy}

Listing~\ref{lst:policy} illustrates a Progent-style~\cite{shi2025progent}
policy rule that mediates filesystem tool access.
The policy is expressed in a declarative domain-specific language rather
than embedded in the LLM prompt, which would ensure that enforcement is
deterministic regardless of planner behavior, illustrating how the
``separate stochastic planning from deterministic execution''
principle (\S\ref{sec:principles:separation}) could be realized.

\begin{figure}[h]
\begin{lstlisting}[
  language={},
  basicstyle=\ttfamily\scriptsize,
  frame=single,
  xleftmargin=2pt,
  xrightmargin=2pt,
  breaklines=true,
  columns=flexible,
  captionpos=b,
  caption={Policy rule gating filesystem access with scoped,
    time-bounded tokens.},
  label={lst:policy}
]
policy "workspace-fs-guard" {

  # Applies to all tools in the fs-workspace server
  match tool.server == "fs-workspace"

  # Layer 2: scope restriction
  require tool.params.path starts_with "./workspace/"
  deny    tool.params.path contains       ".."

  # Layer 2: capability token validation
  require capability.token is_valid
  require capability.ttl   > now()
  require capability.scope includes tool.params.path

  # Layer 2: privilege escalation gate
  if tool.side_effects == "filesystem_mutation" {
    require capability.permissions includes "fs:write"
    if file_size(tool.params.path) > 10MB {
      escalate to user_approval(
        reason: "Large file write detected",
        timeout: 30s,
        default: "deny"
      )
    }
  }

  # Layer 4: audit obligation
  emit audit_event {
    plan_step_id:    context.plan_step,
    tool_server:     tool.server,
    tool_name:       tool.name,
    capability_hash: hash(capability.token),
    decision:        "allow" | "deny" | "escalate",
    timestamp:       now()
  }
}
\end{lstlisting}
\end{figure}

\smallskip
\noindent
Several elements of this example are worth calling out:
\begin{packeditemize}
  \item The \texttt{match} clause binds the policy to a specific tool
    server, avoiding blanket rules that could conflict across servers.
  \item Path validation uses both a positive prefix check and a negative
    traversal check, illustrating defense in depth.
  \item The \texttt{escalate} block demonstrates dynamic privilege
    escalation with a timeout and a default-deny fallback, reflecting
    Progent's context-dependent scope narrowing.
  \item Each policy decision would emit a structured audit event linking
    the decision to the originating plan step and capability token,
    consistent with the Layer~4 provenance patterns.
\end{packeditemize}

\subsection{Memory Provenance Schema Example}
\label{app:examples:memory}

Listing~\ref{lst:memory} shows what a signed memory entry with full provenance
metadata could look like.
This schema illustrates how the ``memory is state'' principle
(\S\ref{sec:principles:memory}) and the Layer~4 integrity patterns
(\S\ref{sec:principles:doctrine:l4}) could be instantiated.
In this design, every persistent memory write would carry a provenance
chain linking the entry back to the originating user intent, the planner
step, the tool invocation that produced the content, and the capability
token under which the write was authorized.
This would make it possible to detect and quarantine entries whose
provenance chain is broken~\cite{sunil2026memorypoisoning} or whose
originating session has been flagged as
compromised~\cite{chen2024agentpoison}.

\begin{figure}[h]
\begin{lstlisting}[
  language={},
  basicstyle=\ttfamily\scriptsize,
  frame=single,
  xleftmargin=2pt,
  xrightmargin=2pt,
  breaklines=true,
  columns=flexible,
  captionpos=b,
  caption={Signed memory entry with provenance metadata.},
  label={lst:memory}
]
{
  "entry_id": "mem-2026-0317-0042",
  "content": "User prefers pytest over unittest for all
              Python projects in this workspace.",
  "content_hash": "sha256:b7e4f1...9c03",
  "provenance": {
    "session_id": "sess-a1b2c3",
    "user_intent": "Configure testing preferences",
    "plan_step_id": "step-007",
    "source_tool": "fs-workspace:read_file",
    "source_invocation_id": "inv-0042",
    "capability_token_hash": "sha256:d4e5f6...1a2b",
    "data_flow_label": "user-preference:trusted",
    "write_authorized_by": "policy:workspace-fs-guard"
  },
  "timestamps": {
    "created": "2026-03-17T14:22:08Z",
    "expires": "2027-03-17T14:22:08Z",
    "last_retrieved": null
  },
  "integrity": {
    "signer": "platform:memory-service",
    "algorithm": "Ed25519",
    "signature": "base64:Mz4wN1...kQ==",
    "certificate_chain": [
      "sigstore://platform/memory-service/2026Q1"
    ]
  },
  "retrieval_policy": {
    "quarantine": false,
    "confidence": "high",
    "max_retrieval_count": null,
    "cross_session_allowed": true
  }
}
\end{lstlisting}
\end{figure}

\noindent
Several fields in this schema carry most of the security burden:
\begin{packeditemize}
  \item The \texttt{content\_hash} would enable tamper detection: any modification to the entry body would invalidate the signature without requiring the verifier to trust the storage backend.
  \item The \texttt{provenance} block would provide end-to-end traceability from memory entry back to user intent, following the in-toto~\cite{torresarias2019intoto} pattern of linking artifacts to their authorized production steps.
  \item The \texttt{data\_flow\_label} records the
    CaMeL-style~\cite{debenedetti2025camel} information-flow label at write time, enabling retrieval-time taint checks that could prevent untrusted tool outputs from silently entering the trusted memory store.
  \item The \texttt{retrieval\_policy} blocks would support operational
    controls: entries could be quarantined, given confidence scores, or
    restricted from cross-session retrieval, addressing the
    cross-temporal persistence threat demonstrated by
    AgentPoison~\cite{chen2024agentpoison}.
  \item The \texttt{expires} field enforces a retention policy, so that stale entries are eventually purged rather than accumulating
    indefinitely.
\end{packeditemize}

\subsection{Audit Telemetry Example}
\label{app:examples:audit}

Listing~\ref{lst:audit} shows what a structured audit log entry for a single
tool invocation could look like, annotated with full provenance.
This schema illustrates how the Layer~4 audit patterns
(\S\ref{sec:principles:doctrine:l4}) and the ``secure agent engineering
is an operations problem'' principle (\S\ref{sec:principles:engineering})
could be operationalized, providing the kind of telemetry substrate
that incident response playbooks, anomaly detection, and post-hoc
forensics would depend on~\cite{dong2024agentops}.

\begin{figure}[h]
\begin{lstlisting}[
  language={},
  basicstyle=\ttfamily\scriptsize,
  frame=single,
  xleftmargin=2pt,
  xrightmargin=2pt,
  breaklines=true,
  columns=flexible,
  captionpos=b,
  caption={Audit telemetry entry for a tool invocation.},
  label={lst:audit}
]
{
  "event_type": "tool_invocation",
  "event_id": "evt-2026-0317-1087",
  "timestamp": "2026-03-17T14:22:09.142Z",
  "session": {
    "session_id": "sess-a1b2c3",
    "user_id": "user-7890",
    "agent_model": "gpt-4o-2025-03",
    "agent_version": "platform-v2.4.1"
  },
  "plan_context": {
    "plan_id": "plan-delta",
    "plan_step_id": "step-007",
    "plan_step_description": "Read pyproject.toml to check
                              test framework configuration",
    "plan_mutation_count": 0,
    "user_intent_hash": "sha256:c3d4e5...f6a7"
  },
  "tool": {
    "server": "fs-workspace",
    "tool_name": "read_file",
    "invocation_id": "inv-0042",
    "parameters": {"path": "./workspace/pyproject.toml"},
    "parameters_hash": "sha256:1a2b3c...4d5e"
  },
  "capability": {
    "token_hash": "sha256:d4e5f6...1a2b",
    "permissions_used": ["fs:read"],
    "scope_validated": true,
    "ttl_remaining_seconds": 2847
  },
  "data_flow": {
    "input_labels": ["user-intent:trusted"],
    "output_labels": ["tool-output:workspace-scoped"],
    "taint_propagation": "read_file -> plan_context.step_008"
  },
  "policy_decision": {
    "policy_name": "workspace-fs-guard",
    "decision": "allow",
    "policy_version": "1.2.0",
    "context_hash": "sha256:e5f6a7...b8c9"
  },
  "outcome": {
    "status": "success",
    "latency_ms": 12,
    "output_hash": "sha256:f6a7b8...c9d0",
    "output_size_bytes": 1847,
    "side_effects": "none"
  }
}
\end{lstlisting}
\end{figure}

\noindent
The most important aspects of this entry are the following:
\begin{packeditemize}
  \item The entry links four provenance dimensions: session identity,
    plan context, tool invocation, and capability token.
    This would enable forensic queries such as ``which tool calls in
    session X used capability token Y after plan mutation Z?''
  \item The \texttt{data\_flow} block would record
    CaMeL-style~\cite{debenedetti2025camel} taint labels for both
    inputs and outputs, plus the downstream propagation target,
    supporting post-hoc taint analysis without re-executing the
    session.
  \item The \texttt{policy\_decision} block would record which policy
    version authorized the invocation and a context hash that enables
    verification of the decision basis, corresponding to the
    audit-breadth metric from the scorecard introduced in
    \S\ref{sec:principles}.
  \item Hashing parameters and outputs rather than logging raw content
    would preserve auditability while respecting data-minimization
    considerations in privacy-sensitive deployments.
  \item The \texttt{plan\_mutation\_count} field tracks how many
    times the plan has been modified since user approval, corresponding
    to the Plan Mutation Coverage metric from the evaluation
    scorecard.
\end{packeditemize}

\end{document}